\documentclass[prl,aps,twocolumn,showpacs,floatfix]{revtex4}
\usepackage{graphicx}
\usepackage{amssymb}
\usepackage{amsmath}
\usepackage{amsfonts}
\usepackage{color}
\usepackage{verbatim}
\usepackage[normalem]{ulem}
\usepackage{lipsum}

\begin{document}

\preprint{APS/123-QED}

\title{
%{\cb Comprehensive Studies of the Bulk Superconducting Order Parameter in Ba(K)Fe$_2$As$_2$}\\

%\cred{Probing Bulk Superconducting Order Parameter in Ba(K)Fe$_2$As$_2$ by Four Complementary Techniques}
Probing Bulk Superconducting Order Parameter in Ba(K)Fe$_2$As$_2$ by Four Complementary Techniques
}

\author{A.~V.~Muratov$^1$} \email{avmuratov@mail.ru}
\author{T. E. Kuzmicheva$^1$}
\author{A.~V.~Sadakov$^1$}
\author{S.~Yu.~Gavrilkin$^1$}
\author{D.~A.~Knyazev$^{1,2}$}
\author{S. A. Kuzmichev$^{1,3}$}
\author{Yu.~A.~Aleshchenko$^{1,4}$}
\author{A. A. Kordyuk $^5$}
\author{V.~M.~Pudalov$^{1,6}$}
\author{M. Abdel-Hafiez$^7$}

%\author{A.~Biankoni$^{1,2,7}$}
\affiliation{$^1$ P. N. Lebedev Physical Institute, Russian Academy of Sciences, Moscow 119991, Russia}
%\affiliation{$^2$ Moscow Institute  of Physics and Technology, Moscow 141700, Russia}
\affiliation{$^2$ International Laboratory of High Magnetic Fields and Low Temperatures, Wroclaw 253-421, Poland}
\affiliation{$^3$ Low Temperature Physics and Superconductivity Department, Physics Faculty, M.V. Lomonosov Moscow State University, 119991 Moscow, Russia}
\affiliation{$^4$ National Research Nuclear University MEPhI (Moscow Engineering Physics Institute), Moscow 115409, Russia}
\affiliation{$^5$ Institute of Metal Physics of National Academy of Sciences of Ukraine, 03142 Kyiv, Ukraine}
\affiliation{$^6$ National Research University Higher School of Economics, 101000 Moscow, Russia}
\affiliation{$^7$ Institute of Physics, Goethe University Frankfurt, 60438 Frankfurt, Germany}
%\affiliation{$^7$ RICMASS, Rome International Center for Materials Science Superstripes,  00185 Rome, Italy}

\date{\today}

\begin{abstract}
Using {\em four} different experimental techniques, we performed comprehensive studies of the {\em bulk} superconductive properties of single crystals of the nearly optimally doped Ba$_{1-x}$K$_x$Fe$_2$As$_2$ ($T_{c} \approx 36$\,K), a typical representative  of the 122 family. We investigated  temperature dependences of the (i) specific heat $C_{el}(T)$, (ii)  lower critical magnetic field $H_{c1}(T)$
%and the anisotropy of the second critical field $H_{c2}$
, (iii) intrinsic multiple Andreev reflection effect (IMARE), and (iv) infrared reflectivity spectra. All data
%are consistent with each other and
clearly show the presence of (at least) two superconducting
%condensates with
nodeless
%order parameters
gaps. The quantitative data on the superconducting spectrum obtained by four different techniques are  consistent with each other:
%{\cg
(a) the small superconducting gap $\Delta_S(0) \approx 1.8 - 2.5$\,meV, and the
%extended $s$-wave
large gap energy $\Delta_L(0) \approx 9.5 - 11.3$\,meV that demonstrates the signature of an extended $s$-wave symmetry ($\sim~33 \%$ in-plane anisotropy),
(b) the characteristic ratio $2\Delta_L/k_BT_C$ noticeably exceeds the BCS value.
%}
%Next,  the ratio of the intraband-to-interband coupling  is an order of 10; the prodominant coupling within each condensate seems to be inconsistent with the $s\pm $ type symmetry.
\end{abstract}

% PACS, the Physics and Astronomy
     % Classification Scheme.

\maketitle

\section{Introduction}
The symmetry structure of Cooper pairs is thought to be the key to the understanding of the pairing mechanism of their superconductivity.
%In iron pnictide superconductors, the precise symmetry of the order parameter
%as well its evolution with doping
%remains highly controversial.
It is well-known that in conventional superconductors the electron-phonon interaction gives rise to the attraction between two electrons, thus forming Cooper pairs. However, Superconductors, whose averaged order parameter over the entire Fermi surface yields zero, are called unconventional. In Iron-based superconductors, the popular opinion is that the electron-phonon is not strong enough to overcome Coulomb repulsion and form Cooper pairs. The nature of the pairing state in iron-based superconductors is the subject of much debate \cite{paglione_NatPhys_2010,Stewart_RMP_2011,Fisher_RPP_2011,Hirschfeld_RMP_2011, Basov_NP_2011, Dai_NP_2012}.

The ternary iron arsenide BaFe$_2$As$_2$ shows superconductivity at about $37-38$\,K by hole doping
 %with partial substitution of potassium, sodium etc. for barium
 \cite{rotter_PRL_2008}. Among various known Fe-based superconductors (FeBS),  these 122 type family compounds  may be grown as high quality and  large size  single crystals with easily variable doping.  Band structure calculations show that the low energy bands are dominated by the
Fe 3$d$ orbitals forming multiple band metallic state: hole-like Fermi surfaces (FS) around the $\Gamma$
(0,0) point and electron-like Fermi sheets around the M $(\pi,\pi)$ point in the Brillouin zone (BZ).
The electron and hole-like FS sheets in the normal state of (Ba$_{1-x}$K$_x$)Fe$_2$As$_2$
%have been
observed in angle-resolved photoemission spectroscopy (ARPES)
%be
are gapped by either the spin density wave (SDW) \cite{yang_PRL_2009, wang_PRL_2009} or superconducting \cite{wu_EPL_2008, chen_EPL_2009} order in the parent ($x=0$)  or superconducting ($x> 0.15$) compound, respectively.

It is well experimentally established that potassium  doping  leads to suppression of the SDW ordering in the parent BaFe$_2$As$_2$ compound and induces superconducting (SC) state. The Hall coefficient and thermoelectric power (TEP) measurements for the parent BaFe$_2$As$_2$  indicate $n$-type carriers, whereas potassium doping leads to change of the sign in Hall and TEP coefficients, thus indicating $p$-type carriers in superconducting Ba$_{1-x}$K$_x$Fe$_2$As$_2$ \cite{wu_EPL_2008}. For the optimal doping $x \approx 0.4$ the superconducting critical temperature reaches $T_c\approx 38$\,K \cite{rotter_PRL_2008, popovich_PRL_2010}.

 In the normal state, the electron and hole sheets of the FS are of a comparable size \cite{zabolotnyy_2009, ding_EPL_2008, zhang_PRL_2010, zhang_NatPhys_2012}.
 %It turns out that the physics of the pairing could be more complicated than it was originally thought, because of the multi-band nature of low-energy electronic excitations.~\cite{chubukov_2012}.
  In the superconducting state, several energy bands at the Fermi energy give rise to multiple energy gaps in the respective superconducting condensates~\cite{paglione_NatPhys_2010}.  Recent specific heat, magnetization, muon spin rotation ($\mu$SR),  tunneling spectroscopy, Andreev reflection spectroscopy, and ARPES measurements provide clear evidence of multiple gap structures in 122-type FeBS.

  The available quantitative  experimental data on the key superconducting parameters  probed by distinct techniques as well as in various experiments are far of being consistent. Also,  identification of the
  %condensates
  superconducting gaps
  with the relevant FS bands  is hampered by the fact that each particular probe is sensitive only in a limited energy range.
 Thus far, thermodynamic specific heat measurements with optimally doped BKFA crystals revealed either two nodeless superconducting gaps,  $\Delta_1 = 11$\,meV and  $\Delta_2  =3.5$\,meV \cite{popovich_PRL_2010}, or one gap: $\Delta=6$\,meV \cite{mu_PRB_2009}, or 6.6\,meV \cite{kant_PRB_2010}. By fitting temperature dependence of the
 %London penetration depth
lower critical magnetic field $H_{c1}(T)$ extracted from low field  magnetization measurements, two superconducting gaps were found in Ref.~\cite{ren_PRL_2008}, $\Delta_1(0)=8.9 \pm 0.4$\,meV, and $\Delta_2(0)=2.0 \pm 0.3$\,meV. Penetration depth extracted from %muon spin rotation measurements (
 $\mu$SR
 leads to  $\Delta_1=9.1$\,meV,  and $\Delta_2=1.5$\,meV \cite{khasanov_PRL_2009}.

 The specific heat (SH) measurements ~\cite{johnson-mahmoud_PRB_2014, pramanik-mahmoud_PRB_2011, popovich_PRL_2010} are known to suffer of several evident problems with data treatment. The SH data contains contribution from the lattice, that is subtracted to some extent in order to determine the electronic contribution. The lattice contribution to the SH is typically estimated by suppressing the SC transition in high magnetic fields or by measuring SH for the parent non-SC compound. Therefore, the lattice SH cannot be accurately obtained in FeBS because of the very high upper critical field and because of magnetic/structural phase transitions at higher temperature in the parent compound. The majority of the earlier SH data suffer from a residual low-temperature non-superconducting electronic contribution and show Schottky anomalies ~\cite{pramanik-mahmoud_PRB_2011, hardy_JPSJ_2014}. Moreover, superconductivity-induced electronic SH is very sensitive to the sample quality and phase purity~\cite{popovich_PRL_2010}. Also, in the earlier SH data analysis, the data are commonly fitted to the phenomenological multiband $\alpha$-model~\cite{padamsee_JLTP_1973,bouquet_EPL_2001}, that assumes a BCS temperature dependence of the gaps. However, our
 %Andreev
 direct measurements by means of multiple Andreev reflections effect (MARE)
 spectroscopy ~\cite{
 kuzmichev_SSC_2012,
 kute-zhigadlo-Sm1111_JETPL_2014,
 kuzmichev-LiFeAs_JETPL_2013,
 kuzmichev-MgB2_JETPL_2014,
 kute-Gd1111_EPL_2013,
 shanygina_JSNM_2013} do not support this assumption and clearly show that the $\Delta(T)$ dependences for the multiband superconductors (such as MgB$_{2}$, and FeBS) deviate from the BCS-type because of the interband coupling. Finally, fitting the SH data with the multiband model requires several adjustable parameters.
  %It might be therefore that a combination of all the above obstacles causes  dissimilar gap values (and, in particular, their unrealistically large values, such as 11 and 3.5 meV~\cite{popovich_PRL_2010}), obtained from the SH measurements.

 The amount of the  SC gaps detected in ARPES measurements varies, depending, apparently,  on the instrument resolution,  crystal and its surface quality:
 % resolution of the ARPES measurements improved progressively:
 initial experiments \cite{ding_EPL_2008} reported large gap $\Delta_1=12$\,meV  on both small hole-like and electron-like FS sheets,  and a small gap $\Delta_2=6$\,meV   on the large hole-like FS;  similar results were reported in  Ref. \cite{khasanov_PRL_2009}: $\Delta_1=9.1$\,meV, and  $\Delta_2 <4$\,meV.
 It should be noted that the small  gaps developed on the inner hole and inner electron FS are difficult to
resolve experimentally in ARPES measurements. Later, some more  SC  nodeless gaps were observed; particularly,  in Ref.~\cite{zhao_CPL_2008} the inner FS sheet around $\Gamma$ point was found to show large ($10 - 12$\,meV) and slightly momentum-dependent gap while the outer FS sheet has nearly isotropic small gap ($7 - 8$\,meV). In  Ref.~\cite{zhang_PRL_2010}
 three hole condensates ($\alpha, \beta ,\gamma$) were found around $\Gamma$ point %(0,0)
 ,  and  one electron condensate ($\eta$) around M-point
 %$(\pi,\pi)$
 of the BZ, all with nodeless  SC gaps. $\Delta_\alpha$ was found warped along $k_z$: $\Delta_\alpha=6-11.5 $\,meV as $11.5\cos(k_xa)\cos(k_yb)+2.1\cos(k_zc)$, whereas $\Delta_\beta$ and $\Delta_\gamma$ were isotropic. The $\eta$ SC-gap is also almost isotropic along $z$ and a rhomb-like anisotropic in the $ab$-plane.  Finally, in STS tunneling measurements two nodeless gaps  $\Delta_2=7.6$, and $\Delta_1 =3.3$\,meV were found in Ref. \cite{shan_PRB_2011}.
  %whereas from Andreev reflection spectroscopy \cite{hafiez-122_PRB_2014} two nodeless gaps were determined, $\Delta_L= $ with $\approx 30\%$ anisotropy, and $\Delta_s=1.7\pm 0.3$\,meV.

Substantial efforts have been made  in order to understand the physics of the pairing mechanism.
 On the theory side, for  the  Fe-based superconductors, which have both electron-like and hole-like pockets, there is general agreement among theoretical approaches \cite{chubukov-efremov_PRB_2008, hirschfeld_PRB_2009, hirschfeld_NJP_2010, chubukov_PRB_2010, bernevig_PRL_2011, platt_PRB_2011} that the starting point to the gap symmetry is the $s^{\pm}$ type with opposite sign of the gap on the electron and hole pockets. This symmetry, however, may change as FS sheets size changes \cite{wu_EPL_2008, castellan_PRL_2011}, or as nonmagnetic impurities are introduced \cite{efremov-korshunov-dolgov_PRB_2011}. The majority of experimental data, cited above reported  nodeless $s$-type symmetry gaps. However, thermodynamic probes are sensitive  to the line nodes  with  sufficiently high spectral weight,  whose existence in BKFA they rule out.
ARPES measurements  with optimally doped (Ba$_{1-x}$K$_x$)Fe$_2$As$_2$ \cite{bernevig_PRL_2011, ding_EPL_2008, nakayama_PRB_2011} and Ba(Fe$_{1-x}$Co$_x$)$_2$As$_2$ \cite{terashima_PNAS_2009}, have  identified  nodeless  gaps  on  the
hole pockets.
%, ruling out a non $s$-wave gap symmetry.
We recall that thermodynamic measurements on these same materials \cite{khasanov_PRL_2009, budko_PRB_2010, prozorov_PRB_2011, hafiez-122_PRB_2014} also show nodeless  behavior consistent with $s$-wave gap symmetry. Recent data of the  $T_c$ dependence on nonmagnetic impurities in Ba(Fe$_{1-x}$Co$_x$)$_2$As$_2$ disordered films \cite{mitsen_JETP_2015}
%seems to be
initially seemed to be
inconsistent with $s^\pm$ type theory predictions \cite{chubukov-efremov_PRB_2008},
%though the pristine polycrystalline films represent already rather disordered material.
however, a more detailed subsequent analysis of the same data \cite{mitsen_2016} lead the authors to the conclusion on  the $s_\pm$ gap symmetry.
Finally, the phase sensitive SIS tunneling measurements \cite{burmistrova_PRB_2015} reported the $s^{\pm}$ symmetry for current injected in the $ab$-plane (and $s^{++}$-wave -- for current injected along $c$).

%On the other hand, the density of states (DOS) calculations show that the states at the Fermi level E$_{F}$ are formed mainly by 3$d$-electrons of Fe, thus the metallic-type conductance is namely due to these 3$d$-states.~\cite{DJS,DK} This leads to the suggestion that any kind of spacer between FeAs blocks affects the level of doping rather than fundamental pairing mechanism. Consequently, one could assume that spacer doping has a minor influence on the superconducting gap symmetry.

We conclude that the existing  experimental data on the gap structure and anisotropy in $k$-space are contradictory enough.
In this context, it is highly important to  probe the superconducting properties with a set of independent experimental techniques. Each of the experimental probes has its own limits of applicability and requires particular model assumptions for extracting the quantitative data from the observables.
Comparing the results obtained by several independent techniques one may test the validity of model assumption and obtain most reliable information.
In Ref. \cite{hafiez-122_PRB_2014} this approach has been implemented  by applying two independent bulk probes, i.e. by measuring  the London penetration depth  and MARE.
%SnS-Andreev reflection spectroscopy.
Despite the fact that well consistent data have been obtained in Ref.~\cite{hafiez-122_PRB_2014} on the gap magnitude, these measurements did not fully address the problem since were performed with similar, though not identical samples and even of the  nominally different composition, Ca$_{0.32}$Na$_{0.68}$Fe$_{2}$As$_{2}$ and Ba$_{0.65}$K$_{0.35}$Fe$_{2}$As$_{2}$.

This drawback is improved in the current study, where we have succeeded in performing  {\em four} types of measurements  with one and the same large size single crystal of nearly optimally doped (Ba$_{1-x}$K$_{x})$Fe$_{2}$As$_{2}$ (with $x=0.33-0.35$).  In particular, we have measured temperature dependences of the specific heat, lower critical field, $H_{c1}$, multiple Andreev reflections effect, and infrared reflectance spectra.
We obtained self-consistent data that clearly shows the presence of two or more superconducting condensates with nodeless order parameters.
%{\cg
The quantitative data on the superconducting properties obtained by four complementary techniques may be summarized as follows: (a) the superconducting state has two (or more) nodeless gaps: the large gap, $\Delta _L = 9.5 - 11.3$\,meV with extended $s$-wave symmetry, and the small gap, $\Delta _S = 1.8-2.5$\,meV; (b) both energy gaps fall with temperature in the way different from the single-band BCS-like  behavior, (c) the characteristic ratio $2\Delta_L/k_BT_C$ noticeably exceeds the BCS limit and indicates rather strong electron-boson coupling in the driving bands.
%The above cited interval for the large gap value includes either two different large gaps (pointing at the existence of three superconducting condensates), or the $\sim 30\%$ anisotropy interval of the large gap.

%Next, given a certain nonzero coulomb repulsion, the ratio of the intraband-to-interband coupling is of the order of 2.4; consequently, the strongly dominant coupling within each condensate seems to be hardly consistent with the $s\pm $ type symmetry for the optimally doped BKFA samples.

%}

%{\cred Add some more on the surface states problem for the surface-sensitive probes (STS, STM, SIS, and overall inconsistency of the data.}

%%%%%%%%%%%%%%%%%%%%%%%%%%%%%%%%%%%%%%%%%%%%%%%%%%%%%%%%%%%%%%%%%%%%%%%%%%%%%%%%%%%%%%%%%%%

\section{Experimental details}

The large size single crystal of Ba$_{1-x}$K$_{x}$Fe$_{2}$As$_{2}$ was synthesized by self-flux technique using FeAs as the flux, for details see ~\cite{shan_NatPhys_2010, luo_SST_2008}.  For Ba-122 FeBS, optimal level corresponds to $x \approx 0.4$ for K doping ($T_c ˜= 38.5$\,K) \cite{popovich_PRL_2010, avci_PRB_2012}.

The chemical composition of our sample was verified by
%scanning electron microscope JEOL JSM-7001 equipped with an
energy dispersive X-ray (EDX) spectroscopy probe.
According to the magnetic susceptibility measurements in zero field (see upper inset of Fig.~\ref{Fig:M(H)}) and specific heat measurements the critical temperature of the superconducting transition $T_c= 36.5 \pm 0.2$\,K. If one relies on the  known phase diagram \cite{avci_PRB_2012}
%the phase composition
the average bulk doping level of the studied samples
 %was refined and
 may  be concluded to correspond to $x=0.33$.

The high quality of the crystals is confirmed by various physical characterizations: (i) a sharp superconducting transition observed in susceptibility and specific heat measurements  at $T\approx 37$\,K~\cite{johnson-mahmoud_PRB_2014} (see inset of Fig.\ref{Fig:M(H)})
%(ii) the large value of the residual resistivity ratio (RRR) is {\cred found to be $\rho_{(300\,K)}$/$\rho_{(36\,K)}$ = 12.8???? - ask Tanya}
confirming the good quality of the single crystal~\cite{johnson-mahmoud_PRB_2014}; (ii) the chemical composition,  crystal structure and lattice parameters tested by X-ray diffraction (Pan Analytical X'Pert Pro MRD). The critical temperature $T_{c} \approx$ 36.5-37\,K,  is evidenced by magnetization, DC transport measurements, and also by Andreev reflection spectra flattening measured at various points of the bulk crystal.

Low field magnetization measurements were performed by using a  SQUID magnetometer MPMS-XL7, and specific heat measurements - with PPMS-9 system, both  from Quantum Design. Infrared reflectance (IR) spectra were measured with IFS-125HR Fourier transform infrared spectrometer from Bruker, and  Andreev reflection spectra were obtained by the break-junction technique \cite{kute-Gd1111_EPL_2013, kute-uspekhi_2014}.

%\begin{figure}
%\includegraphics[width=240pt]{xi(T).eps}
%\caption{Magnetic susceptibility of the Ba$_{0.67}$K$_0.33$Fe$_2$As$_2$ %sample,{\cred measured in field ????? Plot the opj-figure. Add the field %cooling data.}}
%\label{Fig:xi(T)}
%\end{figure}

\section{DC magnetization}
\label{DC_magnetization}
The London penetration depth $\lambda$ is a fundamental parameter that carries  signatures of the pairing mechanism, and therefore is a powerful tool for probing the superconducting state~\cite{prozorov_RPP_2011}.
%In order to determine  $\lambda$ and its temperature dependence,  we performed precise measurements of the first critical magnetic field   $H_{c1}(T)$,
The London penetration depth is related to lower critical field $H_{c1}$, that pinpoints the  vortices penetration into the sample.

In this section we report measurements of the first critical field $H_{\rm c1}$ for Ba$_{1-x}$K$_x$Fe$_2$As$_2$ sample.
Our analysis
%shows that the superconducting gaps determined through fitting to the
 %London penetration depth
of temperature dependence of the lower critical field $H_{c1}(T)$ for the $B|| c$ direction support  the presence of two $s$-wave-like gaps with strongly different magnitudes and slightly different  contributions.
By analyzing the $H_{\rm c1}$ temperature dependence we reveal the presence  of the two SC condensates with $s-$type symmetry of the order parameter.
%{\cred
The two SC gap values extracted from the $H_{\rm c1}(T)$ analysis correspond to
$2\Delta_1(0)/k_BT_c =1.2 \pm 0.2$ (or $\Delta_1(0)= 2 \pm 0.3$\,meV), and $2\Delta_2(0)/k_BT_c = 6.9 \pm 0.3$ (or $\Delta_2(0)= 11 \pm 0.5$\,meV); their weights extracted by fitting with the $\alpha-$model  are  $\varphi_1=0.54 \pm 0.02$ for the small gap and $\varphi_2=0.46 \pm 0.02$ for the large gap.
%}

\subsection{Experimental}
The DC magnetization measurements were performed with a rectangular slab, $3\times 4 \times 0.05$ mm$^3$, cleft from the same large crystal used for all other measurements.
%The critical temperature determined from magnetic susceptibility in zero field and from specific heat equals to $T_c= 36.5 \pm 0.2$\,K.
%In this section we report measurements of the first critical field $H_{\rm c1}$ for Ba$_{1-x}$K$_x$Fe$_2$As$_2$ sample.
%From the critical superconducting temperature  $T_c=36.5$\,K we identified the average bulk doping level $x=0.33$.

The approach used for extracting the first critical magnetic field is based on measuring the magnetic field value, for which the vortexes start penetrating into superconducting bulk destroying the ideal Meissner effect. In other words,  we determined such field value $H_{\rm c1}$ which corresponds to  the onset of nonlinear $M$ versus $H$ dependence. Measurements were performed with MPMS-XL7 (Quantum Design) in the temperature range  $2-36$\,K with a step size of 1\,K. Magnetic field direction was aligned with the crystal $c$-axis.

\begin{figure}
\includegraphics[width=250pt]{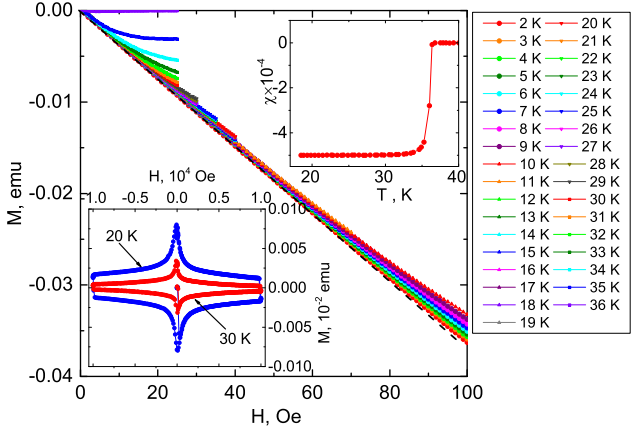}
\caption{Magnetic field dependence of the magnetization in the temperature range 2--36\,K. The dashed straight line extrapolates the linear $M(H)$ dependence observed in weak fields. Upper inset shows magnetic susceptibility of the Ba$_{0.67}$K$_{0.33}$Fe$_2$As$_2$ sample, measured in zero field. Lower inset: magnetization hysteresis loops measured at 20 and 30 K.   }
%zooms in the $M(H)$ dependence for several temperatures in the vicinity of $T_c$.}
\label{Fig:M(H)}
\end{figure}

\subsection{Results}

At first step we checked pinning properties by measuring of magnetic hysteresis loops at several temperatures (lower inset to Fig.~\ref{Fig:M(H)}). Magnitization curve $M(H)$ is symmetric about the axis M=0 that indicates a strong bulk pinning and the absence of Bean-Levingston barier. Also $M(H)$ shows no magnetic background.

The raw experimental data for $M(H)$ dependences in low fields are presented in Fig.~\ref{Fig:M(H)}.
In fields above $H_{c1}$ the superconductor captures magnetic flux, that leads to departure of $M(H)$ dependence from the linear one.
Exact finding of the $H_{c1}$ values from the measured nonlinear $M(H)$ dependence is a hard task, taking into account a finite width of the linear-to-nonlinear crossover of the magnetization curves, and data scattering.
In our measurements the noise level corresponded to $\approx (3-5)\times 10^{-5}$\,emu. By modeling  the $M(H)$ dependence with such noise level we found that the frequently used algorithm for the $H_{c1}$ determination based on the correlation parameter, (see e.g., \cite{hafiez-122_PRB_2014, hafiez-FeSe_PRB_2013})
%appeared to be inapplicable.
leads to  artificially overestimated $H_{c1}$ data and excessive
%large
$H_{c1}(T)$  data scattering.

\begin{figure}
\begin{center}
\includegraphics[width=240pt]{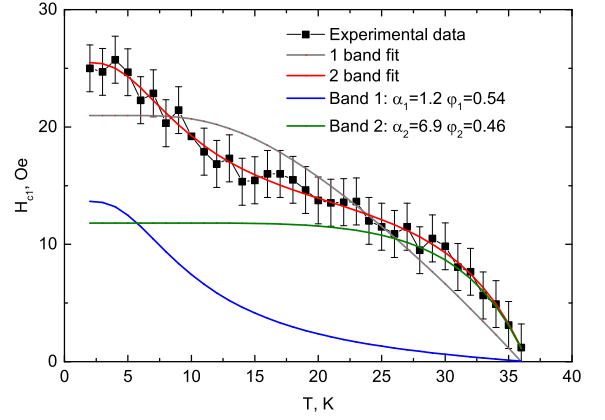}
\caption{$H_{c1}$ temperature dependence for the BKFA sample compared with the single band BCS model and the two band $\alpha$-model}
\label{Fig:Hc1}
\end{center}
\end{figure}

Correspondingly, instead of  the above algorithm \cite{hafiez-122_PRB_2014}  based on regression calculation, we have developed a modified algorithm  where the experimental $M(H)$ data  are fitted with
%a joint using of
both, linear (for  $H<H_{c1}$) and the second power polynomials (for $H>H_{c1}$).  The  protocol of Refs.~\cite{hafiez-122_PRB_2014, hafiez-FeSe_PRB_2013} and  its shortcomings in the case of a large noise level, as well as the modified algorithm are described  in detail in Appendix 1.
This approach minimizes  the impact of a variable number of points on the correlation coefficient calculation  and thus improves the accuracy of the $H_{c1}$ determination.
The  $H_{c1}(M)$ dependence determined with the modified algorithm for the studied BKFA sample is shown in Fig.~\ref{Fig:Hc1}

It should be mentioned that the determined $H_{c1}$ value
%dependence
represents a critical field for the given sample.
In order to characterize the material parameter,  one has also to take the demagnetization factor $N$ into account:
\begin{equation}
H_{c1}= \frac{H_{c1}^{\rm measured}}{1-N},
\end{equation}
where
\begin{equation}
N=\frac{q\frac{a}{b}}{q\frac{a}{b}+1}.
\end{equation}
For the disk-shape sample \cite{brandt_PRB_1999}:
\begin{equation}
q_{\rm disk}=\frac{4}{3\pi} +\frac{2}{3\pi}
\tanh \left[1.27\frac{b}{a} \left( \ln\left(1+\frac{a}{b}\right)\right) \right].
\end{equation}
With the sample diameter $a=3\pm0.5$\,mm and thickness $b=50\pm 20$ $\mu$m the demagnetization factor for our sample $N = 0.96 \pm 0.02$, and
the ratio of the material $H_{c1}$  and the measured $H_{c1}$  value amounts to $\sim 16-50$.
Correspondingly, the  $H_{c1}$ value for Ba$_{0.67}$K$_{0.33}$Fe$_2$As$_2$  falls into a range of fields $400-1250$\,Oe.

%\subsection{Results: $H_{c1}(T)$ and its analysis}

For describing the lower critical field of a superconductor, it is convenient to introduce a normalized superfluid density \cite{ren_PRL_2008}:
\begin{equation}
\tilde{\rho_s}(T)\equiv\lambda(0)_{ab}^2/\lambda(T)_{ab}^2=H_{c1}(T)/H_{c1}(0).
\end{equation}

In the framework of the BCS theory for a single-band superconductor with an isotropic gap, the latter may be represented as
\cite{carrington_2003, luo_PRL_2005}:
\begin{equation}
\tilde{\rho_s}(T) = 1+2\int_{\Delta(T)}^\infty \frac{\partial f}{\partial E} \frac{EdE}{\sqrt{E^2-\Delta^2(T)}}.
\end{equation}
Here $f(T)$ is the Fermi distribution function, $\Delta(T)=\Delta_0
%\delta(T)$  and $\delta(T)=
\tanh\left[1.82 \left(1.018 \left(\frac{T_c}{T}-1\right)\right)^{0.51} \right]$ is the gap temperature dependence.
$E^2=\varepsilon^2+\Delta^2 (T)$, $E$ is the total energy, and $\varepsilon$ - single particle excitation energy counted from the Fermi energy.
The normalized superfluid density may be re-written in  a more convenient for integration way as follows:
\begin{equation}
\tilde{\rho_s}(T) = 1-2 \int_0^\infty \frac{e^y}{(e^y+1)^2}\frac{1}{t}d\xi,
\end{equation}
with $t=T/T_c$,
%\begin{eqnarray}
$y=\sqrt{\xi^2+\left(\frac{\alpha \delta(T)}{2}\right)^2}/t$,
%\\nonumber\\
$\xi=\varepsilon/\left(k_B T_c\right) $,
%\nonumber \\
and $\alpha = 2\Delta_0/\left(k_B T_c \right)$.
%\alpha &=& \frac{2\Delta_0}{k_B T_c}.
%\end{eqnarray}
The latter is the parameter in the given model.

Figure \ref{Fig:Hc1} shows the least square fitting of the measured $H_{c1}(T)$ data within the above single-band BCS model. The fitting parameter here $\alpha=3.4$. One can see that the model fails to reproduce the experimental data. Clearly, the single band model can not describe the curved $H_{c1}(T)$ dependence, especially in the interval $10-25$\,K.
%\begin{figure}
%\begin{center}
%%\includegraphics[width=240pt]{Hc1(T)-single_band.eps}
%\caption{$H_{c1}$ temperature dependence for the BKFA sample compared with the single band BCS model}
%\label{Fig:Hc1-single band}
%\end{center}
%\end{figure}
The physical meaning of this failure is transparent: to fit the data successfully one needs to use a multiband model.

Correspondingly,
at the next step for describing the experimental data we apply the  so called two-band $\alpha$-model \cite{carrington_2003}:
\begin{equation}
\tilde{\rho_s} (T)= \varphi_1 \tilde{\rho}_{s1}(T)+\varphi_2 \tilde{\rho}_{s2} (T)
\end{equation}

%\begin{figure}
%\begin{center}
%\includegraphics[width=240pt]{Hc1(T)-two_band.eps}
%\caption{$H_{c1}$ temperature dependence for the BKFA sample compared with the two band $\alpha$ model. Red curve %- full normalized superfluid condensate density, green and blue curves are the contributions from the  first and %the second bands {\cred Combine the three figures Hc1(T)} }
%\label{Fig:Hc1-two_band}
%\end{center}
%\end{figure}

This model considers a normalized superfluid density for the superconductor having two independent  condensates  with a normalized superfluid densities $\rho_{s1}$ and $\rho_{s2}$ in the first and second band respectively, taken with weighting factors $\varphi_1$ and  $\varphi_2=1-\varphi_1$.

%\subsection{Fitting results}
The result of fitting  the $H_{c1}(T)$ data with $\alpha$ model is shown in Fig.~\ref{Fig:Hc1}. This approach leads to a good agreement between the model and experimental data.
%{\cred
The fitting parameters are as follows:
$\alpha_1 = 1.2 \pm 0.2$ ($\Delta_1(0)= 2 \pm 0.3$\,meV), weight factor $\varphi_1=0.54 \pm 0.02$; and
$\alpha_2= 6.9 \pm 0.3$ ($\Delta_2(0)= 11 \pm 0.5$\,meV), weight factor $\varphi_2=0.46 \pm 0.02$;
 $H_{c1}(0) = 25.5$\,Oe.
%}

\section{Specific heat}

 The specific heat measurements are a powerful thermodynamic bulk probe
  \cite{johnson-mahmoud_PRB_2014, pramanik-mahmoud_PRB_2011, popovich_PRL_2010}, though there are several known problems with SH data treatment.
 %, however.
 The SH data contains contribution from the lattice, that is to be subtracted  in order to determine the electronic SH. The lattice contribution to the SH is usually estimated by suppressing the superconducting transition in high magnetic fields. For FeBS, the lattice SH cannot be suppressed because of the very high upper critical field. The majority of the earlier SH data suffer from a residual low-temperature non-superconducting electronic contribution and show a Schottky anomaly ~\cite{pramanik-mahmoud_PRB_2011, hardy_JPSJ_2014}.
%Moreover, superconductivity-induced electronic SH is very sensitive to the sample quality and phase purity~\cite{popovich_PRL_2010}.

In this section we report our SH data and their analysis which evidence for the two-band superconducting condensate with $s-$type order parameter symmetry. The extracted superconducting gap values correspond to the characteristic ratios  $2\Delta_1(0)/k_BT_c =1.6 \pm 0.1$ ($\Delta_1(0)= 2.5 \pm 0.2$\,meV, weight factor $\varphi_1=0.58 \pm 0.02$) and $2\Delta_2(0)/k_BT_c = 7.2 \pm 0.2 (\Delta_2(0)= 11.3 \pm 0.3$\,meV, weighting factor $\varphi_2=0.42 \pm 0.02$). These parameters are consistent with those determined from magnetization measurements, IR reflection, and Andreev reflection spectra, descried in the corresponding sections.

\begin{figure}[h]
\begin{center}
\includegraphics[width=240pt]{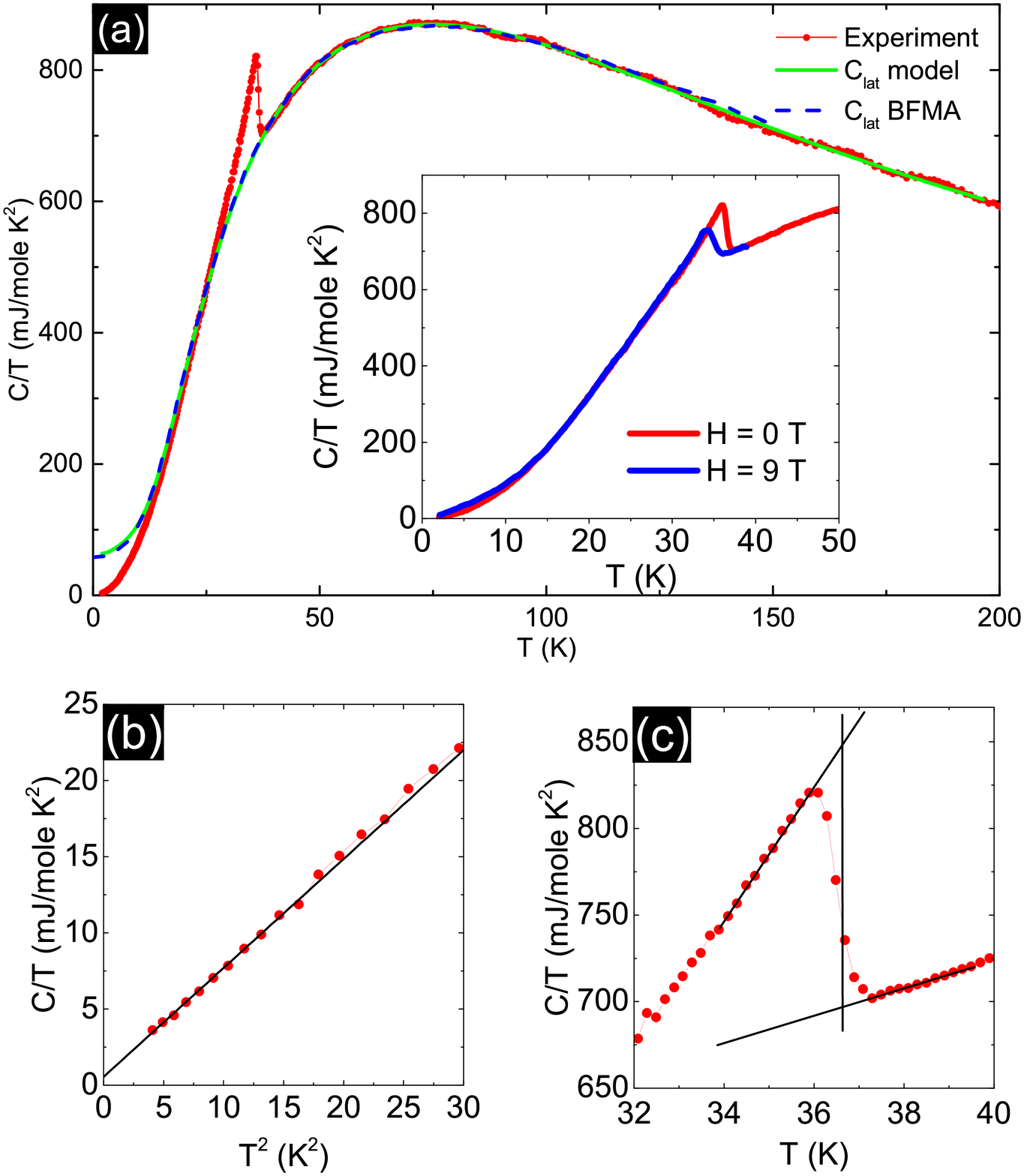}
\caption{(a)Temperature dependence of the specific heat for Ba$_{0.67}$K$_{0.33}$Fe$_2$As$_2$ sample, normalized by temperature at zero field and comparison with two models, using the common states approximation: for the 6-modes Einstein  model and for the lattice SH of  Ba(Fe$_{0.88}$Mn$_{0.12}$)$_2$As$_2$. Insert compares the data for $B=0$ and $B=9$\,T.  (b) Low-temperature behavior of the normalized specific heat  versus $T^2$. Dots are the experimental data, black curve $-$ their fit with the Debye law. For quantitative extraction of the parameters we used only the data in the range  $2-4$\,K.   (c) SH anomaly at the superconducting transition and  determination of $T_c$.  }
\label{Fig:C(T)}
\end{center}
\end{figure}

% \subsection{Experimental details}
\subsection{Experimental}
 The specific heat  measurements were taken with a 1.93\,mg-piece of Ba$_{1-x}$K$_x$Fe$_2$As$_2$ ($x=0.33$) single crystal  cleft from the same large crystal that was used for all other measurements; the sample had superconducting critical temperature $T_c=36.5$\,K.
Measurements were done using the thermal relaxation technique with  PPMS-9, in the temperature range $2 - 200$\,K. Temperature was swept with a stepsize of  0.2\,K for the interval  2 - 50\,K, 0.5 K for 50 - 100\,K and 1\,K for 100 - 200\,K. For each temperature point the data have been averaged within 3 seconds.

\subsection{Results}

The raw experimental SH data are shown in Fig.~\ref{Fig:C(T)} at zero field; the insert shows results obtained with another piece of the same crystal  (m=0.8\,mg) in fields $B=0T$ and $9T$.

The $C(T)/T$ data shows no features in the low temperature range  (such as, \textit{e.g.}, growth towards the lowest $T$), thus evidencing for the absence of Schottky anomaly. In the low-$T$ limit (for $T < 6-8$\,K) the $C(T)/T$ data may be represented by the Debye law: $C(T)/T=\gamma(0)+\beta T^2$, where $\gamma(0)$ is the  residual contribution of the non-superconducting phase, $\beta T^2$ - is the lattice contribution. The two parameters $\gamma(0)$ and $\beta$ may be easily found from fitting the model to the experimental data (see Fig.~\ref{Fig:C(T)}\,b). For $B= 0$ we found $\gamma= 0.3 - 0.5$ mJ/mol K$^2$ and $\beta= 0.74 - 0.71$\,mJ/mol K$^4$
%{\cred (the parameters spread is determined by the amount of the experimental points????)}.
The  negligeably low value of the residual  electronic specific heat evidences for high quality of the sample. It is worth noting that the above approach is rather approximate because beyond the linear approximation the electronic SH for superconducting materials depends on temperature, and  because the lattice contribution includes higher order terms. For this reason this approach is appropriate only for qualitative estimates, whereas  for quantitative analysis more complex approach is needed, which is described below.

%\begin{figure}
%\begin{center}
%\includegraphics[width=240pt]{LowT-C(T).eps}
%\caption{Low-temperature behavior of the normalized specific heat  versus $T^2$. Dots are the experimental data, %black curve - their fit with Debye low. For quantitative extraction of the parameters we used only the data in the %range  $2-4$\,K.  {\cred Plot the opj-figure}}
%\label{Fig:LowT-C(T)}
%\end{center}
%\end{figure}

%\subsection{Specific heat data and their processing}
In the temperature interval 36-37\,K the $C(T)$ data shows a sharp peak, related with the SC transition (see Fig.~\ref{Fig:C(T)}). The peak width is about  1\,K, and the jump in the $C/T$ data  at the transition  $\Delta C/T = 119$\,mJ/mol K$^2$.
%From the peak position in Fig.~\ref{Fig:C(T)}\,c we estimate critical superconducting temperature $T_c = 36.5$\,K almost coinciding with the value determined  from magnetic measurements.
%In order to determine $T_c$ we linearly extrapolated the experimental data, as shown in Fig.~.
Due to the entropy conservation at the SC transition, the following equality must be fulfilled:
\begin{equation}
\int_{T_c-t}^{T_c+t}\frac{C_{exp}}{T} dT=\int_{T_c-t}^{T_c}\frac{C_{\rm extrap}}{T}dT+\int_{T_c}^{T_c+t}\frac{C_{\rm extrap}}{T} dT,
\end{equation}
where $C_{exp}$ - the measured SH data, $C_{\rm extrap}$ - the data extrapolated to the region of the SC transition, and $2t$ is the superconducting transition width.
By implementing this implicit equation  to the data in Fig.~\ref{Fig:C(T)}\,c we determine the true $T_c$ value of 36.5\,K, nicely consistent with that extracted from magnetic measurements.

%\begin{figure}
%\begin{center}
%\includegraphics[width=170pt]{DeltaC.eps}
%\caption{SH anomaly at the superconducting transition.  {\cred Plot the opj-figure.}}
%\label{Fig:DeltaC}
%\end{center}
%\end{figure}

%\subsubsection{Separation of the lattice and electronic contributions to SH}
\subsubsection{Separation of the lattice and electronic contributions to SH}
Further experimental investigations of the structure and magnitude of the SC gaps  by means of bulk specific heat data are of great interest. In order to determine the specific heat related to the SC phase transition, we need to estimate the phonon (lattice) and electronic contributions to specific heat in the normal state.
%The key and most hard task in the SH data analysis  is the separation of electronic and lattice contributions.
These contributions are additive:
\begin{equation}
C_p(T)=C_e(T)+C_{lat}(T),
\end{equation}
where $C_e$ is the contribution related to electronic subsystem, and $C_{lat}$ is the lattice contribution. The lattice term $C_{lat}$ however cannot be determined by direct measurements.
This problem may be solved  by using the so called common states approximation  \cite{stout_1995}, that consists in using, as a reference, of the lattice SH for a non-superconducting compound of a
%neighboring
relative's composition. For Ba$_{0.67}$K$_{0.33}$Fe$_2$As$_2$ one of such
%neighboring
%sister
compounds is the parent BaFe$_2$As$_2$ that is non-superconducting though exhibits a magnetic phase transition at $\approx 140$\,K. Varying doping level or doping element leads to changes in the lattice spacings by a few percents. In order to take account of these insignificant change one can use scaling factors proximate to unity. Other possible
%examples of the
reference materials for our Ba$_{0.67}$K$_{0.33}$Fe$_2$As$_2$ sample are the nonsuperconducting  compounds Ba(Fe$_{0.85}$Co$_{0.15}$)$_2$As$_2$ \cite{hardy_PRB_2010}, Ba(Fe$_{0.88}$Mn$_{0.12}$)$_2$As$_2$  \cite{popovich_PRL_2010}, and BaFe$_{1.75}$Ni$_{0.25}$As$_2$ \cite{hafiez_1501.01655}.

Mathematically, the  common states approximation for the specific heat  may be written as follows:
\begin{equation}
C_{\rm tot}^{SC}(T)=C_{el}^{SC}(T) + A C_{\rm lat}^{nSC}(BT).
\end{equation}

Here $C_{\rm tot}^{SC}(T)$ is the total calculated SH corresponding to the experimental data $C_{\rm exp}(T)$, $C_{el}^{SC}(T)$ is the electronic contribution to SH, $C_{\rm lat}^{nSC}(T)$ - lattice SH for the nonmagnetic
%intimate
reference compound, $A$ and $B$ are the scaling factors.
For temperatures above $T_c$ the electronic SH   may be written as  $C_{el}^{SC}(T) = \gamma_n T$.
The  factors $A$  and $B$ are selected based on least square  fitting  under the constraint of the entropy conservation:
\begin{equation}
 \int_0^{T_c}\frac{C_{el}(T)}{T} dT =  \int_0^{T_c}\gamma_n dT.
\end{equation}

By now, the lattice specific heat for Ba$_{0.68}$K$_{0.32}$Fe$_2$As$_2$ was well described using the Debye-Einstein model \cite{popovich-SOM_PRL_2010}.
In order to test whether or not we can apply the data of Refs.~\cite{popovich_PRL_2010, popovich-SOM_PRL_2010, hafiez_1501.01655} to the data processing for our Ba$_{0.67}$K$_{0.33}$Fe$_2$As$_2$ sample, we have tested the results of Ref.~\cite{popovich_PRL_2010, popovich-SOM_PRL_2010, hafiez_1501.01655} for the lattice  SH of non-superconducting and non-magnetic materials, and found that these data may be scaled to each other by using the common state approximation with factors  $A$ and  $B$ chosen rather close to unity, $0.95 - 1.05$.

\begin{figure}
\begin{center}
\includegraphics[width=240pt]{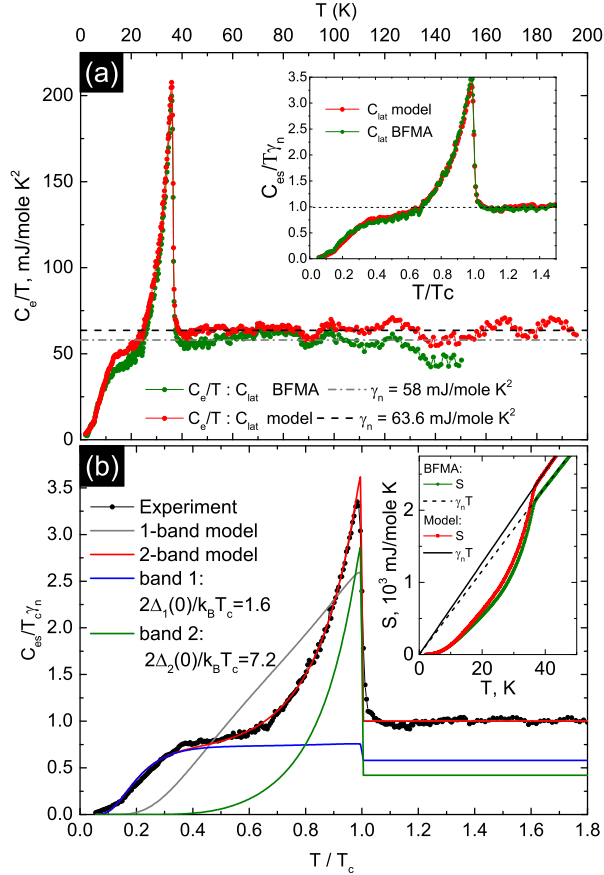}
\caption{(a) Electronic SH calculated using the 6-mode Einstein model \cite{popovich-SOM_PRL_2010} and the lattice SH of Ba(Fe$_{0.88}$Mn$_{0.12}$)$_2$As$_2$ \cite{popovich_PRL_2010}.  Inset: comparison of the normalized electronic SH defined by the 6-mode Einstein model and the lattice SH of Ba(Fe$_{0.88}$Mn$_{0.12}$)$_2$As$_2$; (b) Normalized electronic SH of the superconducting condensate $C_{es}/T \gamma_n$ compared with the single band BCS model and the two band BCS $\alpha$-model. Inset: Electronic entropy in normal and SC state in case of calculations with 6-mode Einstein model and the lattice SH of Ba(Fe$_{0.88}$Mn$_{0.12}$)$_2$As$_2$.}
\label{Fig:CeN}
\end{center}
\end{figure}

Correspondingly, for the analysis of our experimental data we used the model described in Ref.~\cite{popovich-SOM_PRL_2010} containing 6 Einstein modes. We also used the lattice SH data for Ba(Fe$_{0.88}$Mn$_{0.12}$)$_2$As$_2$ \cite{popovich_PRL_2010}, since these measurements were done in the most wide temperature range. Figure~\ref{Fig:C(T)}\,a  shows that both models describe  the experimental $C(T)$ data rather well. The resulting electronic SH contribution $C_{el}/T$ obtained in this fit using the common state approximation
%and entropy conservation constrain,
is shown on Fig.~\ref{Fig:CeN}\,a, the inset to Fig.~\ref{Fig:CeN}\,b demonstrates the entropy conservation constraint for this calculations.
In the two respective fittings we obtained the two sets of factors:  (i) $A= 0.998$, $B= 0.974$ and  $\gamma_n= 63.6$\,mJ/mol\,K$^2$ for the 1st scaling based on the lattice SH of Ba$_{0.68}$K$_{0.32}$Fe$_2$As$_2$, and (ii)  $A= 0.954$, $B= 0.996$,  $\gamma_n= 58.0$\,mJ/mol\,Ê$^2$ for the 2nd scaling based on the lattice SH of Ba(Fe$_{0.88}$Mn$_{0.12}$)$_2$As$_2$.

In order to improve the accuracy of the data analysis we extrapolated the $C_{el}(T)$ data to $T= 0$ for both models.  From this extrapolation we obtained also the two estimates for normal state residual contribution, $\gamma_r=2.3$\,mJ/mol\,K$^2$ - in the  analysis based on the lattice model \cite{popovich-SOM_PRL_2010} and $\gamma_r = 1.1$mJ/mol\,K$^2$  in the scaling based on the Ba(Fe$_{0.88}$Mn$_{0.12}$)$_2$As$_2$ lattice. We conclude that the non-superconducting residual contribution to SH is of the order of $2-4$\% ($\gamma_r/\gamma_n=0.019- 0.035$), that is  comparable to the values reported for other superconducting FeBS \cite{hardy_PRB_2010}.

Analyzing behavior of the superconducting condensate it is appropriate to consider the normalized electronic SH $C_{es}/T\gamma_n$ versus normalized temperature $(T/T_c)$, where $C_{es}$ is the SH of superconducting condensate \cite{hardy_PRB_2010}, which may be obtained as:
\begin{equation}
\frac{C_{es}}{\gamma_nT}(T)=\frac{C_e/T-\gamma_r}{\gamma_n-\gamma_r}
\end{equation}

In Fig.~\ref{Fig:CeN}\,a, the $C_e/T$ data
%fitting
obtained
by two approaches is consistent with each other. Although there is a minor difference  (much less than the peak height at $T_c$) between them  in Fig.~\ref{Fig:CeN}\,a,  the difference becomes almost invisible on the plot of the normalized SH of the superconducting condensate $C_{es}$, Fig.~\ref{Fig:CeN}\,b. For high temperatures,
$T > 100$\,K, the data description based on the  Ba(Fe$_{0.88}$Mn$_{0.12}$)$_2$As$_2$ lattice SH is somewhat worse: the difference between  $C_e/T$ and è $\gamma_n$  increases with temperature.

%\subsubsection{Analysis of the normalized electronic SH}
\subsubsection{Analysis of the normalized electronic SH}
The normalized SH of the superconducting condensate may be calculated within the BCS theory as follows \cite{bouquet_EPL_2001}:
\begin{eqnarray}
\frac{C(T)}{\gamma_nT} &=&\frac{d(S/\gamma_nT_c)}{dt}, \nonumber \\
\frac{S(T)}{\gamma_nT_c} &=& \frac{6}{\pi^2} \int_0^\infty \left[f\ln f +(1-f)\ln(1-f) \right]d\varepsilon, \nonumber \\
f &=&  \left[\exp\frac{\left(e^2+\alpha^2\delta^2(t)/4\right)^{1/2}}{t} +1  \right]^{-1}, \nonumber \\
\delta(T) &=& \tanh\left[ 1.82\left( 1.018\left(\frac{T}{T_c}-1  \right) \right)^{0.51}\right].
\label{Eq:BCS-SH}
\end{eqnarray}
where $t=T/T_c$, $\alpha= 2\Delta(0)/k_BT_c$, $\delta(T) =\Delta(T)/\Delta(0)$ is the temperature dependence of the gap,  and  $\Delta(0)$  is the energy gap at $T= 0$.
The above phenomenological formulae \cite{hafiez-122_PRB_2014} generalizes  calculations of  \cite{muhlschlegel_1959} within the BCS model.

At the first step, for fitting the $C_{es}(T)$ data we applied the single-band BCS model using Eqs.~\ref{Eq:BCS-SH}. The model implies an isotropic $s$-type order parameter $\Delta$. Figure~\ref{Fig:CeN}\,b shows the result of the mean square fitting with  $2\Delta/k_BT_c= 3.7$. Obviously, the single-band approach does not fit the experimental SH data and, particularly, does not reproduce the remarkable hump in $ C_{es}/T\gamma_n$ clearly seen
at  $T/T_c \sim 0.3-0.5$.

At the second step we apply the phenomenological $\alpha-$model \cite{bouquet_EPL_2001, padamsee_JLTP_1973}  for the  two-band superconductor, which sums up contributions of each band, calculated within the BCS model,  Eq.~\ref{Eq:BCS-SH}, with the corresponding weight factors $\varphi_1$ and $\varphi_2= 1-\varphi_1$:
\begin{equation}
C(T)= \varphi_1 C_1(T) + \varphi_2 C_2(T)
\end{equation}

This model has three adjustable parameters,  $\alpha_1=2\Delta_1/k_BT_c$, $\alpha_2=2\Delta_2/k_BT_c$ and $\varphi_1$, which may be found from least square fitting of the model to the experimental data. $\varphi_1$ and $\varphi_2$ describe the relative share of each condensate in the total SH: $\varphi_i= \gamma_i/\gamma_n$, where $\gamma_i$ is the specific heat of the $i$-th condensate in the normal state. The result of data fitting with the two-band model is shown in Fig.~\ref{Fig:CeN}\,b.

One can see that the two-band approach provides rather good fitting to the experimental data. The difference between the model dependence and the experimental data does not exceed  5\%  of $C_{es}/T\gamma_n$, that corresponds to  4\,mJ/mol\,K$^2$. The deviation is within the measurements uncertainty and in relative units does not exceed  1\% of the total measured $C_{\rm exp}$. With the two band model we find the following set of parameters:
%{\cred
$\alpha_1=2\Delta_1/k_BT_c = 1.6 \pm 0.1$ ($\Delta_1= 2.5 \pm 0.2$\,meV), $\alpha_2=2\Delta_2/k_BT_c = 7.2 \pm 0.2$ ($\Delta_2 = 11.3 \pm 0.3$\,meV), and $\varphi_1=0.58 \pm 0.02$.
%}
%%%%%%%%%%%%%%%%%%%%%%%%%%%%%%%%%%%%%%%%%%%%%%%%%%%%%%%%%%%%%%%%%%%%%%%%%%%%%

\section{Infrared reflection spectroscopy}

Infrared (IR) spectroscopy is a powerful technique to investigate the electronic gap structure of superconductors.  Its large probe depth ensures the bulk nature of the measured quantities and its high-energy resolution and powerful sum rules enable a reliable determination of important physical parameters, such as the gap magnitude and the plasma frequency of the SC condensate \cite{basov-2005}. In a simple one-band system, the standard Drude model with parameters plasma frequency $\Omega $ and scattering rate $\gamma $ describes the frequency-dependent complex conductivity $\tilde{\sigma }_N$ in the normal (N) state \cite{burns-1990}. In the superconducting (S) state, the standard BCS model (Mattis-Bardeen equations \cite{mattis-1958}, with parameters $\sigma _0$  and superconducting gap $\Delta $) can describe the complex conductivity $\tilde{\sigma}_S$ \cite{tinkham-1975}. On this basis, far-infrared measurements can be of particular importance since a signature of the superconducting gap $\Delta $) can be observed at $\hbar\omega\sim 2\Delta$ (optical gap) for an anisotropic $s$-wave BCS superconductor. The electromagnetic radiation below the gap energy $2\Delta $ could not be absorbed. For a bulk sample, in particular, a maximum at the optical gap is expected in the ratio $R_S/R_N$, where $R_S$ and $R_N$ are the frequency-dependent reflectances in the superconducting and normal state, respectively \cite{tinkham-1975}.

\subsection{Experimental}
IR reflectance spectra $R(\omega )$ were measured with Bruker IFS 125HR spectrometer with a spectral resolution of 2 cm$^{-1}$ over a wave number range of 400-50 cm$^{-1}$ (25-200 $\mu $m).
For measurements in FIR region  we used a mylar beam splitters of various thickness. Liquid-helium cooled Si bolometer was used to detect IR spectra. For low-temperature measurements the sample was placed into the helium cryostat Optistat CF-V from Oxford Instruments with the wedged windows made of TPX plastic. The reflectance measurements were carried out at near-normal incidence on the freshly cleaved surfaces.

The goal of IR measurements is to determine the frequency-dependent complex conductivity $\tilde{\sigma }(\omega )=\sigma _1(\omega ) + i\sigma _2(\omega )$ which usually appears in discussion of the low-frequency electrodynamics of the system \cite{basov-2005} and can describe its optical response. The complex conductivity of the ideal single-band conducting system can be described using the Drude model in the normal state and the Bardeen-Mattis BCS model \cite{mattis-1958} generalized for an isotropic $s$-wave BCS superconductor using the Zimmermann relations \cite{zimmermann-1991}. In this case, in the "dirty" limit the dissipative part of the optical conductivity $\sigma _1(\omega )$ at $T\ll T_c$ vanishes abruptly below a frequency corresponding to doubled superconducting gap $2\Delta $. Thus, in the vicinity of the frequency corresponding to $2\Delta $ (optical gap) one should observe a peculiarity in the optical response of the system.

The relatively small size of the sample for IR measurements and irregular cleavage surface resulted in rather low accuracy of measurement of the absolute value of reflection coefficient; the latter hampered calculating the optical conductivity by using the  Kramers-Kronig analysis. For this reason we apply the technique described in \cite{palmer-1968} to determine the superconducting gaps. It consists in the relative measurements of $R(T\ll T_c)/R_N$ with no reference measurements, while sweeping  the temperature within a narrow temperature range. Here, $R_N$ is the reflectance in the normal state at temperature slightly above $T_c$. The measurements are performed in one cycle with the same detector and set of optical elements (beam splitter and cryostat windows). In this way the sample position and orientation as well as the optical system were not changed during measurements. This technique enables to minimize possible temperature-driven distortions of the optical set-up, which may yield frequency-dependent systematic errors in $R(\omega )$. It should be noted that for the bulk superconductor of the $s-$type symmetry the normalized reflectivity $R(T\ll T_c)/R_N$ forms a maximum, whose energy corresponds to the superconducting gap $2\Delta $. For the two gap superconductor, the maximum is expected to appear between the  two SC gaps, closer to the one having a major contribution. This enables one to estimate the value of the dominant gap.

\subsection{Results}
\begin{figure}
\includegraphics[width=250pt]{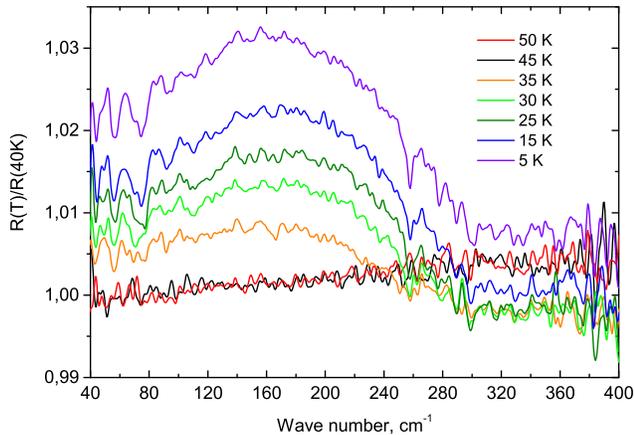}
\caption{$R(T)/R$(40\, K) dependences of Ba$_{0.67}$K$_{0.33}$Fe$_2$As$_2$ measured within the temperature range $T=5-50$\,K.}
\label{Fig:IR}
\end{figure}

Figure~\ref{Fig:IR} shows the $R(T)/R(T=40$\,K)
dependences for Ba$_{0.67}$K$_{0.33}$Fe$_2$As$_2$ measured at $T=5-50$\,K. One can see that the normalized reflectivity $R(T)/R(40$\,K) maximum starts increasing as temperature $T$ decreases below $T_c$. This is because for $s$-wave superconductor at temperatures below $T_c$ the reflectance approaches unity at energies $\hbar\omega < 2\Delta $. As a result, a peak is formed with a maximum in the range of $\sim 160$ cm$^{-1}$ (19.8 meV). The peak position correlates with the magnitude of the greater of the superconducting gaps \cite{popovich_PRL_2010, evtushinsky-2009, charnukha-2011, shan_PRB_2011, evtushinsky-2014}. The smaller gap is beyond the frequency range of our IR measurements. The kink in the normalized reflectivity at $\sim 250$ cm$^{-1}$ is probably due to the IR active phonon mode $E_u$ related to the Fe($ab$)-As($-ab$) vibrations \cite{Schafgans-2011}. This mode manifests itself in many AFe$_2$As$_2$ materials including A = Ca, Sr, Eu and Ba.

%%%%%%%%%%%%%%%%%%%%%%%%%%%%%%%%%%%%%%%%%%%%%%%%%%%%%%%%%%%%%%%%%%%%%%%%%%%%%%%%%%%%%%

\section{Intrinsic multiple Andreev reflection effect (IMARE) spectroscopy}
 %{\cg
  In ballistic mode, superconductor - normal metal - superconductor (SnS) contact (whose diameter $2a$ is less than the carrier mean free path $l$ \cite{Sharvin}) demonstrates multiple Andreev reflection effect (MARE) \cite{OTBK,Arnold,Averin,Kummel}. MARE manifests itself in an excess current at low bias voltages in current-voltage characteristic (CVC) of SnS contact (so called foot area). A series of dynamic conductance features called subharmonic gap structure (SGS) appears at bias voltages $V_{n} = 2\Delta/en$ (where $n$ is a natural number) \cite{OTBK,Arnold,Averin,Kummel,Devereaux}. This simple formula enables to directly determine the superconducting gap value at any temperatures
  %$T = 0 -$
  up to $T_c$ \cite{OTBK,Kummel}. For the high-transparency SnS-Andreev regime (typical for our break-junction contacts), SGS exhibits a series of dips for both nodeless and nodal gap \cite{Devereaux,Cuevas,BJ}. The coexistence of two independent superconducting gaps would cause, obviously, two SGS's in the $dI(V)/dV$-spectrum. The $k$-space angular distribution of the gap value strongly affects the SGS lineshape. In case of an isotropic gap, the SGS minima are high-intensive and symmetrical, whereas a nodal gap (such as  $d$-wave) leads to strongly suppressed and asymmetric $dI(V)/dV$ minima \cite{Devereaux, Cuevas,BJ}. For extended $s$-wave nodeless symmetry, the SGS demonstrates doublet minima corresponding to the gap extremes in the $k$-space \cite{kuzmichev-LiFeAs_JETPL_2013,BJ}.

\subsection{Experimental}

For Andreev spectroscopy studies, we used a break-junction technique (for details, see \cite{Moreland, BJ}) in order to create symmetric SnS contacts. The studied sample is precisely cracked in cryogenic environment. We cut from the single crystal a thin plate, $3 \times 1.5 \times 0.1$\,mm$^{3}$. The crystal was attached to a springy holder by four In-Ga pads which insured true 4-probe connection and helped aligning the $ab$-plane parallel to the sample holder. After cooling down to 4.2\,K, the sample holder was precisely bent, which caused cracking of the single crystal. Its deformation generates a microcrack that represents the superconductor - constriction  - superconductor contact (ScS), where the constriction
%acts as a weak link.
formally acts as insulator or normal metal.
In our setup, the superconducting banks are kept touching each other and not separated to a valuable distance \cite{BJ}. Taking in mind the metallic-type Ba spacers between superconducting Fe-As blocks of crystal structure, a formation of a metallic-type constriction is feasibly. The observed $I(V)$ and $dI(V)/dV$ of the break junctions are typical for high-transparent SnS-Andreev mode \cite{OTBK,Arnold,Averin,Kummel}. Obviously, a current flows through the break junction along the $c$-direction (for the details see \cite{BJ}), therefore, a gap anisotropy could be barely resolved in $k_xk_y$ plane \cite{BJ}. Since in our setup the microcrack is located deep in the bulk of the sample and away from current leads, the cryogenic clefts are free of Joule overheating, and adverse surface influence such as possible degradation or impurity diffusing.

In layered sample, the break-junction probe often shows also array of the SnSn-\dots-S-type realized in natural steps and terraces onto cryogenic clefts of layered crystal. In such arrays, an intrinsic multiple Andreev reflections effect
%(IMARE)
 occurs. This effect is similar to the intrinsic Josephson effect \cite{PonIJE, Nakamura} and was first observed in Bi cuprates \cite{PonIMARE}, further in all layered superconductors (\cite{kute-Gd1111_EPL_2013}, for a review, see \cite{BJ}). Since Andreev array consists of a sequence of $m$ identical SnS-junctions, the SGS dips appear at positions:

\begin{equation}
V_{n} = \frac{m \times 2\Delta_i}{en}, m = 1, 2, \dots
\label{Eq:Vnm}
\end{equation}

In case of stack contacts, positions of other peculiarities caused by {\em bulk} properties of material also scale by a factor of $m$ \cite{kute-Gd1111_EPL_2013,BJ}. In our experiment, we were able to probe tens of
%single SnS-contacts and
arrays (containing various number of junctions $m$) by precisely readjusting the microcrack. The latter opportunity helps one to collect a large amount of data and to check reproducibility of the {\em bulk} gap values and other peculiarities caused by {\em bulk} properties of material. The number of junctions $m$ can be determined by normalizing the spectrum of array to that of the single SnS-contact; after such scaling, positions of each SGS should coincide. Probing such $natural$ stack contacts, one obtains information about the true bulk properties of the sample (almost unaffected by surface states which seem to be significant in  Ba-122~\cite{van-heumen_2011} {\em locally}, i.e. within the contact size $a \approx 7-50$\,nm. This feature favors accuracy increasing in the superconducting gap measurements~\cite{kute-Gd1111_EPL_2013}.

\subsection{Results}
\begin{figure}
\includegraphics[width=20pc,clip]{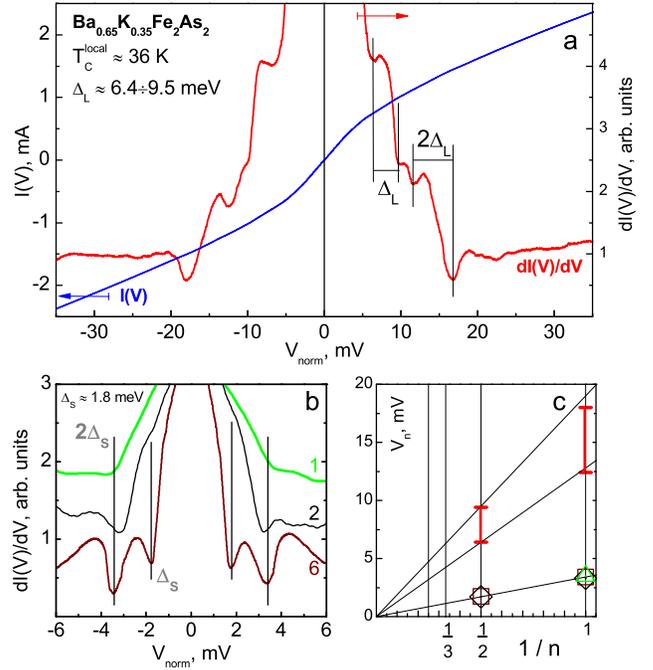}
\caption{a) Normalized to a single junction current-voltage characteristic (left axis) and dynamic conductance spectrum
(right axis) for Andreev array in Ba$_{0.65}$K$_{0.35}$Fe$_2$As$_2$ with $T_c^{local} \approx 36$\,K measured at $T=4.2$\,K.
Black vertical lines depict the positions of subharmonic gap structure (SGS) dips, corresponding to $\Delta_L \approx 6.4 \textendash 9.5$\,meV ($~ 33\%$
anisotropy in k-space angle distribution). b) Low-bias fragments of dynamic conductance spectra demonstrating SGS (marked with black vertical lines) of the small gap $\Delta_S \approx 1.8$\,meV. The $dI(V)/dV$ of $m=6$ and $m=2$ contacts (lower curves) were normalized to the single SnS-junction spectrum (upper curve). Monotonic background was suppressed for clarity. c) The positions of $\Delta_L$ (vertical bars depict the gap anisotropy) and $\Delta_S$ (open symbols) subharmonics in the dynamic conductance spectra shown in a,b panels. Black lines are guidelines.}
\label{Fig:IMARE}
\end{figure}

Figure \ref{Fig:IMARE}a shows a typical current-voltage characteristic (blue line, $T=4.2$\,K) for a break-junction in nearly optimal Ba$_{0.65}$K$_{0.35}$Fe$_2$As$_2$ with critical temperature $T_c^{\rm local} \approx 36$\,K. The excess current at low bias voltages (foot area) manifests a formally metallic-type constriction with ballistic $c$-axis transport \cite{OTBK,Arnold,Averin,Kummel}. Taking the contact resistance $R \approx 15$\,{\rm $\Omega$}, the bulk in-plane resistivity of the studied crystal $\rho^{ab} \approx 0.4 \cdot 10^{-5}$\,{$\rm \Omega \cdot cm$}, and using the value $\rho^{ab}l \approx 0.45 \cdot 10^{-9}$\,{$\rm \Omega \cdot cm^2$} \cite{Zverev}, we estimate the elastic mean free path of carriers $l^{ab} \approx 1.1$\,{$\rm \mu m$}, and the contact radius $a = \sqrt{\frac{4}{3\pi} \cdot \frac{\rho^{ab}l}{R}} \approx 36$ nm.
%$\,{$\rm \mu m$}.
This rough estimation gives the contact dimension
%$2a$ nearly twice less than the carrier mean free path $l$, thus evidencing a possibility of MARE %observation and $2-3$ Andreev subharmonics visible in dynamic conductance spectrum \cite{Kummel}.
$2a << l$, which satisfies the conditions of MARE observation.

The corresponding $dI(V)/dV$ spectrum (red line in Fig.~\ref{Fig:IMARE}a) shows a set of dynamic conductance dips typical for clean classical SnS-Andreev array of 2 junctions (a natural SnSnS structure). In order to normalize $I(V)$ and $dI(V)/dV$ in Fig.~\ref{Fig:IMARE}a to those for a single SnS-junction, the voltage axis was scaled by a factor of $m=2$. The large gap SGS starts with the clear dips at $\approx \pm 18$\,mV corresponding, in accordance with the SGS expression, to $2\Delta_L/e$. The next features at $\approx \pm 11.5$\,mV do not match the expected position ($\approx \pm 9$\,mV) of the second subharmonic of the large gap, therefore these two dips could be interpreted as a doublet $n =1$ feature caused by a $~ 33~\%$
gap anisotropy. The positions of the next pair of dynamic conductance features, $\approx \pm 9.6$ and $\approx \pm 6.4$\,mV, corresponds well with those of the second subharmonic of the large gap. Note that the $n=2$ doublet is right twice narrower than the $n=1$ one, agreeing with the subharmonic set. To say, whether the $dI(V)/dV$ doublets are caused by the in-plane gap anisotropy in the $k$-space, or the order parameter fine splitting,
%in the real space (with a coexistence of two close gaps),
one needs a further study of the dynamic conductance lineshape. In Fig.\ref{Fig:IMARE}a, the real shape of $\Delta_L$ subharmonics is rather ambiguous since overlapped by the pronounced excess conductance. Surely, the intensity and the shape of the $\Delta_L$ dips is inconsistent with that expected for $d$-wave or fully anisotropic (nodal) $s$-wave symmetry \cite{Cuevas, Devereaux}; we conclude therefore that the large gap is nodeless. On the other hand, comparing the current data with those obtained earlier with Ba(K)-122 single crystals with a bit lower $T_c \approx 34$\,K \cite{hafiez-122_PRB_2014}, the extended $s$-wave symmetry of the large gap is more likely.

Using Eq.~(15), we directly determine the large gap edges $\Delta_L^{min} \approx 6.4$\,meV, $\Delta_L^{max} \approx 9.5$\,meV, and corresponding BCS ratios $2\Delta_L/k_BT_c \approx 4.1 - 6.1$. When trying to regard this array as corresponding to a single SnS-junction, we get the twice BCS-ratio up to 12.2 seemed too large for Ba-122; on the other hand, given $m \geq 3$, we get $2\Delta_L/k_BT_c \approx 2 - 3$ which is impossible for driving gap since lies below the weak-coupling limit 3.5. This simple check demonstrates a way for correct determination of the number of junctions in the array; in the case, the 2-junction structure is identified unambiguously.

Figure \ref{Fig:IMARE}b shows low-bias fragments of dynamic conductance spectra of $m=6$ and $m=2$ Andreev arrays, and a single SnS junction (upper curve). The width and the outlook of the pronounced foot near zero bias is reproducible in all the curves. The monotonic background was suppressed in order to clarify the small gap SGS. Black vertical lines in Fig. \ref{Fig:IMARE}b mark the first feature at $V_{S1} \approx \pm3.5$\,mV and the second feature at $V_{S2} \approx \pm1.8$\,mV. These subharmonics, obviously, do not belong to the large gap SGS (see Fig. \ref{Fig:IMARE}a), rather, they originate from a small gap $\Delta_S \approx 1.8$\,meV. Unlike the $\Delta_L$ dips, the small gap peculiarities are not split and are rather symmetric, thus pointing to nearly isotropic $\Delta_S$ in $k-$space. Despite the fact that the three dynamic conductance spectra shown in Fig.~\ref{Fig:IMARE}b are obtained with different Ba(K)-122 samples (with the same $T_c$), the positions of $\Delta_S$ SGS's are reproducible. The sharpening of Andreev features with the $m$ increasing is a representative for IMARE spectroscopy \cite{kute-Gd1111_EPL_2013,BJ} and evidences the bulk nature of the $\Delta_S$ order parameter.

The dependence of SGS positions $V_n$ $versus$ their inverse number $1/n$ shown in Fig.~\ref{Fig:IMARE}c agrees with Eq.~\ref{Eq:Vnm} and represents straight lines crossing the origin. Two independent SGS observed in $dI(V)/dV$ spectra are caused by a presence of at least two distinct condensates with $\Delta_L$ and $\Delta_S$ order parameters.

\begin{figure}
\includegraphics[width=20pc,clip]{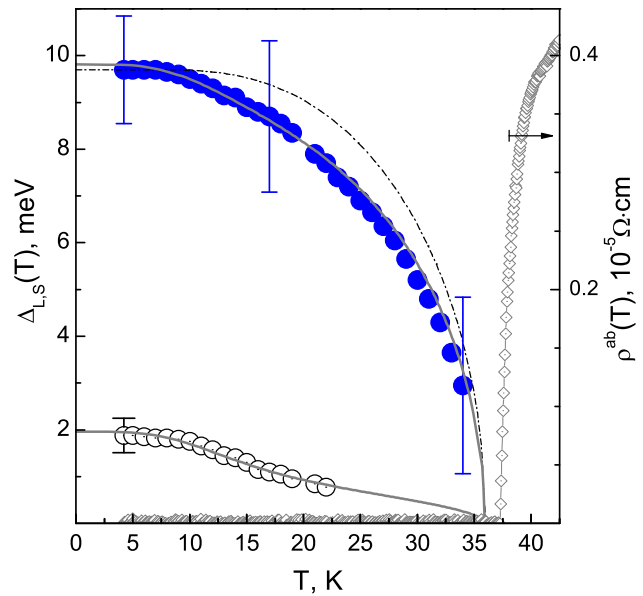}
\caption{Temperature dependences of the large gap (blue solid circles) and the small gap (open circles) in Ba$_{0.65}$K$_{0.35}$Fe$_2$As$_2$. Single-band BCS-like curve (dash-dot line) and bulk resistive transition (rhombs) are shown for comparison.}
\label{Fig:gapt}
\end{figure}

The temperature dependences $\Delta_L(T)$ (corresponding to the positions of the outer dip of doublet-like SGS) and $\Delta_S(T)$ obtained directly are shown in Fig.~\ref{Fig:gapt}. The dependence of the inner $\Delta_L$ extremum is an issue of further studies. The local critical temperature (corresponding to the contact area of $~ 1$\,{\rm $\mu$m} size transition to the normal state) $T_c^{local} \approx 36$\,K is a bit lower than the bulk $T_c$ determined with a bulk probe (see the resistive transition in Fig.~\ref{Fig:gapt}). A single-band model (dash-dot line), obviously, is inconsistent to describe the experimental temperature dependences of the large and the small gaps. $\Delta_L(T)$ passes below the single-band BCS-like curve, whereas $\Delta_S(T)$ bends down significantly. These deviation from the single-band type are caused by a moderate interband interaction. As a result, both gaps turn to zero at common critical temperature $T_c^{local}$.

To approximate the experimental $\Delta_{L,S}(T)$, we used a two effective bands model based on Moskalenko and Suhl gap equations \cite{Mosk,Suhl} with a renormalized BCS-integral. The shape of gap temperature behavior depends on a set of electron-boson coupling constants $\lambda_{ij} = V_{ij}N_j$, where $i,j = L,S$, $V_{ij}$ are matrix interaction elements, $N_{i,j}$ \textemdash normal density of states (DOS) in the corresponding bands at the Fermi level. We took the Debye energy $\hbar\omega_D = 20$\,meV \cite{Rettig}; as fitting parameters, we used $\alpha = N_S / N_L$ ratio (hereafter ``L'' index is linked with the driving bands), and the relation between intra- and interband coupling $\beta = \sqrt{V_LV_S}/V_{LS}$, the fitting is detailed in \cite{kuzmichev-MgB2_JETPL_2014,fit2016}. Theoretical $\Delta_{L,S}(T)$ shown by solid lines agree well with the experimental dependences, therefore, the simple two effective bands model is applicable to describe the IMARE data. The observed $\Delta_{L,S}(T)$ are typical for a strong intraband coupling in the driving bands. The large gap BCS-ratio
%of the $\Delta_L$
far exceeding the weak-coupling BCS limit, also favors the latter statement. In contrast, the Moskalenko-Suhl fit proves a weak-pairing superconductivity in the driven bands solely. In a hypothetical case of zero interband interaction ($V_{LS} = 0$), we estimate $2\Delta_S/k_BT_c^S \approx 3.5$ ($T_c^S$ is the eigen critical temperature of the bands where the small gap is developed).

Taking zero Coulomb pseudopotentials $\mu^{\ast}=0$ suggested, for example, in \cite{Maiti,Mazin}, we get $\lambda_{LL} = 0.59$,  $\lambda_{SS} = 0.28$, $\lambda_{LS} = 0.24$, $\lambda_{SL} = 0.02$ leading to extremely large DOS ratio $\alpha\sim$ 12, and intra- to interband coupling ratio $\beta \sim 6$, which is impossible for the so-called s$^{\pm}$ scenario proposed in \cite{Maiti,Mazin}. When accepting a moderate nonzero Coulomb repulsion $\mu^{\ast}=0.2$, we roughly estimate $\lambda_{LL} = 0.73$,  $\lambda_{SS} = 0.43$, $\lambda_{LS} = 0.4$, $\lambda_{SL} = 0.13$, seemed more realistic. In the latter case, $\alpha \approx 3$, whereas the intraband coupling is $~ 2.4$ times stronger than interband one.
%}

%but this ratio is two times higher than in Ba$_{0.65}$Na$_{0.35}$Fe$_{2}$As$_{2}$.~\cite{pramanik-mahmoud_PRB_2011}
%%%%%%%%%%%%%%%%%%%%%%%%%%%%%%%%%%%%%%%%%%%%%%%%%%%%%%%%%%%%%%%%%%%%%%%%%%%%%%%%%%%%%%%%%%%
\section{Discussion}

\begin{table*}
\caption{\label{tab:table 1}  {The superconducting transition temperature $T_{c}$, and the superconducting gap properties extracted from
%IMARE and lower critical field ($H_{c1}$)
%studies %for
data obtained for Ba$_{1-x}$K$_x$Fe$_{2}$As$_{2}$ by different methods.}}
\begin{ruledtabular}
\begin{tabular}{cccccccccc}
Compounds &$T_c$(K) & Nodes,     & $\Delta_L$(meV) & weight$_L$& $\Delta_S$(meV) &weight$_S$& $\Delta_L/\Delta_S$ &  Technique & Ref.\\
          &         &  anisotropy& $(2\Delta_L/k_BT_c)$&           &$(2\Delta_S/k_BT_c)$&          &                     &            &    \\
\hline

$x=0.33$ &36.5     &no                 &$11.0\pm 0.5$&0.46 & $2.0\pm 0.3$& 0.54  & $5.5\pm 0.3$  &magnetization & this work\\
          &         &  & $(6.9\pm0.3)$&           &$(1.2\pm0.2)$&          &                     &            &    \\
\hline

$x=0.33$ &36.5     &no                &$11.3 \pm 0.3$& 0.42 & $2.5\pm 0.2$ & 0.58  & $4.52 \pm 0.3$ & specific heat & this work\\
          &         &  & $(7.2\pm0.2)$&           &$(1.6\pm0.1)$&          &                     &            &    \\
\hline

$x=0.33$ &36.5     & --              & $9.9 \pm 1$     & -- &  --           & --   & --              & IR normalized  & this work\\
          &         &  & $(6.3\pm0.6)$&           &     &          &                     &            &    \\
\hline

$x=0.35$ & $36 \pm 1$ & no, $\approx 33$\% & $\Delta_L^{min} = 6.4 \pm 0.7$,  &  --   & $1.8 \pm 0.3$ &  --& 3.6 -- 5.3  &     break-junction & this work\\
          &           &                   & $\Delta_L^{max} = 9.5 \pm 1.0$  &     &               &  &      &                    &             \\
          &         &   & $(4-6\pm0.6)$&           &$(1.1\pm0.2)$&          &                     &            &    \\
\hline

$x=0.4$  &35.8     &no                 &$8.8\pm 0.3$ & 0.3& $2.2\pm 0.2$ & 0.7  & $4\pm 0.3$  &magnetization &~\cite{ren_PRL_2008}\\
          &         &   & $(5.7\pm0.2)$&           &$(1.4\pm0.1)$&          &                     &            &    \\
\hline

$x=0.32$ &38.5     &no                &$11$& 0.5 & $3.5$ & 0.5  & $3.14$ & specific heat & ~\cite{popovich_PRL_2010}\\
          &         &   & $(6.6)$&           &$(2.2)$&          &                     &            &    \\
\hline

$x=0.35$&34$\pm 3$&no, $\approx 30$\% & 5.8--8.0  & -- &  $1.7\pm 0.3$ & --  & $4.4\pm 0.3$  & break-junction & \cite{hafiez-122_PRB_2014} \\
          &         &   & $(4-5.5)$&           &$(1.2\pm0.2)$&          &                     &            &    \\
\hline

$x=0.32$ & 38.5    & no              & $10$             & -- & $3.7$            & --   &2.7               & IR          & \cite{charnukha-2011}\\
          &         &   & $(6)$&           &$(2.2)$&          &                     &            &    \\
\hline

$x=0.4$ & 38    & no              & $10\pm1$             & -- & $(3-5.5)\pm1$            & --   &               & ARPES          & \cite{evtushinsky-2014}\\
          &         &   & $(6.1\pm0.6)$&           &$(1.8-3.4\pm0.6)$&          &                     &            &    \\
\end{tabular}
\end{ruledtabular}
\end{table*}

%{\cg
The gap values obtained using the four complementary techniques are summarized in the Table 1. $H_{c1}$ and $C(T)$ probe bulk properties, IR spectroscopy provides information about crystal subsurface layer, whereas IMARE is a direct local probe of the bulk order parameter.
%}
Our experimental data $C_{el}(T)$ and $H_{c1}(T)$ may be well fitted with the two isotropic nodeless gaps. The Andreev spectroscopy data
%squeezes this choice: it
points at two distinct gaps, the anisotropic large gap and isotropic small gap.
%{\cg
 All the data converge on the absence of nodes for both gaps. For the large gap, we report the BCS ratio $2\Delta_L/k_BT_c = 6.1 - 7.2$ exceeding the BCS-limit. This slight variation could be caused by several reasons, such as (a) out-of-plane anisotropy of the order parameter discussed in \cite{Saito}, (b) a complex and nontrivial in-plane angle distribution of the large gap in the $k$-space, (c) a possible presence of a
 %third gap
 large gap splitting, (c) a surface sensitivity of superconducting properties, (d) a significant contribution of high-energy ($\omega > \Delta$)
 pairs with $Re[\Delta(\omega)] > \Delta_{exp}$ (where $\Delta_{exp}$ is a gap edge of the Eliashberg function) accounted in bulk probes. As for the small gap, the determined values give $2\Delta_S/k_BT_c = 1.2 - 1.6$ which lies well below the 3.5 limit and point to a nonzero interaction between the condensates.
%}

% For the anisotropic $s$-wave, the fitting  with the magnitude of the gap $\Delta _0$ = 1.86\,meV is shown in Fig.???? with an anisotropy parameter $\approx$1.09.
%As can be seen the anisotropic $s$-wave order parameter presents a well description to the data.
%(iv) Equally good description of the experimental data for the two-gaps $s$-wave model is obtained using values of $\Delta _{1}(0)$ = 2.26\,meV, $\Delta _{2}(0)$ = 7.28\,meV.
%The gap values for each gap are shown individually in Fig.\,4(d).
It is noteworthy that our extracted gap values are comparable with the two-band $s$-wave fit, $\Delta _{1,2}(0)$= 2 and 8.9\,meV, reported for Ba$_{0.6}$Ka$_{0.4}$Fe$_{2}$As$_{2}$
in ~\cite{ren_PRL_2008}
and $\Delta _{1,2}(0)$= 3.5 and 11\,meV in ~\cite{popovich_PRL_2010}.
%\textcolor[rgb]{1.00,0.00,0.00}{
The value of the gap amplitudes obtained for this material scales relatively well with its $T_{c}$ in light of the recent results for the FeBS~\cite{hafiez-122_PRB_2014, ponomarev_JSNM_2013}.
%}
%In addition, one can notice that the extracted ratio for anisotropic $s$-wave order parameter is smaller than the BCS value, which points to the existence of the large gap.

%{\cb
It is important to note that ARPES studies also report two $s$-wave nodeless gaps of 2.3 and 7.8 meV for the outer and the inner Fermi surface sheets, respectively ~\cite{evtushinsky_PRB_2013}. In fact, ARPES results hint towards the conclusion about strong dependence of the gap value on orbital character of the bands forming the corresponding Fermi surfaces: the larger gap appears on d$_{xz}$/d$_{yz}$ bands~\cite{kordyuk_2012}. Very recently, and based on a multi-band Eliashberg analysis, for Ca$_{0.32}$Na$_{0.68}$Fe$_{2}$As$_{2}$ the superconducting electronic specific heat was shown to be described by a three-band model with an unconventional $s_{\pm}$ pairing symmetry with gap magnitudes of approximately 2.35, 7.48, and formally -7.50 meV~\cite{johnson-mahmoud_PRB_2014}. It has been well demonstrated that the model based on Eliashberg equations is a simplified model of the real four bands model taking into account the similarities between the two 3D Fermi sheets and between the two 2D Fermi sheets. Based on them for the determination of $T_{c}$ and for the gap functions there can be considered only a distinct gap for every 2D, and respectively 3D sets of bands~\cite{dolgov_2005}. In fact,
%in order to solve the Eliashberg equations, there were two ways.
the Eliashberg equations may be solved in two ways.
The first way is to solve the equations which contain dependences of real frequency, and the second one -- to solve this equations on the imaginary axis, summing on Matsubara frequencies~\cite{scalapino_1966}. Thus, the uncertainty in the number of SC condensates to be involved into the data processing affects the parameters extracted from the experiment. In this work we used the simple $\alpha$-model that is not self-consistent, but
%provides a popular model with which
is  often used by
experimentalists
%can fit
for fitting
their thermodynamic data that deviate from the BCS predictions and
%to quantify
for quantifying
those deviations ~\cite{johnston_2013}. From the temperature dependence of the lower critical field data or specific heat data alone it is difficult to be sure whether one, two or three bands can describe well our investigated system, since in the case of multiband superconductivity low-energy quasiparticle excitations can be always explained by the contribution from an electron group with a small gap.

By complementing presented data as well as the data on BaFe$_{1.9}$Ni$_{0.1}$As$_2$ single crystals  ($T_c \approx $19 K) \cite{JSNM2016_Stripes} obtained with MARE spectroscopy with the existing ARPES results \cite{ding_EPL_2008,evtushinsky_PRB_2013,evtushinsky-2014,khasanov_PRL_2009}, one could make conclusion on ab-plane anisotropy of the large order parameter $\Delta_L$.
%However, these data does not exclude the existence of three gaps in the condensate, where the two large gaps different by $\sim 30\%$ mimic the $k-$ space anisotropy, and the third one has an isotropic gap.
%By complementing these data with Andreev reflection spectroscopy we can make more definite conclusion on the structure of SC condensate in BKFA, though still have two possibilities: (i) either two gaps with anisotropic large gap and isotropic small gap, or (ii) three gaps with two comparable large gaps, and a small gap.
%}
%{\cg
%To somehow resolve this issue, it is reasonable to compare the IMARE data with literature. Similar in-plane anisotropy of the large gap, and $2\Delta_{L,S}/k_BT_c$ were obtained in our early IMARE probes with nearly optimal BKFA with $T_c \approx 34$\,K \cite{hafiez-122_PRB_2014} and BaFe$_{1.9}$Ni$_{0.1}$As$_2$ single crystals with $T_c \approx 19$\,K \cite{JSNM2016_Stripes}. Both gaps roughly scale with $T_C$. In this sense, for the driving gap, turning from in-plane anisotropy into two $s$-wave gaps seems unrealistic, since otherwise would mean a dramatic change of underlying pairing mechanism with a minor $T_c$ variation.
%Favoring the latter, the existing ARPES data \cite{ding_EPL_2008,evtushinsky_PRB_2013,evtushinsky-2014,khasanov_PRL_2009} reported two distinct amplitudes of superconducting gaps barely, whereas a third gap was not resolved.
Comparing the $H_{c1}$, $C(T)$, and IMARE data, the two-band model seems to be sufficient to describe the experimental temperature dependences of superconducting parameters.
%}

%%%%%%%%%%%%%%%%%%%%%%%%%%%%%%%%%%%%%%%%%%%%%%%%%%%%%%%%%%%%%%%%%%%%%%%%%%%%%%%%%%%%%%%%%%%

\section{Conclusions}

%{\cg
Using four complementary experimental techniques, we studied single crystals of  the 122 family, nearly optimally doped Ba$_{1-x}$K$_x$Fe$_{2}$As$_{2}$,  and obtained consistent data on the structure of the superconducting order parameter. Our  data extracted from (i)  temperature dependence of
%the penetration depth $\lambda^{-2}_{ab}(T)$
lower critical field,  and (ii) temperature
%and field
dependence of the specific heat,  are  inconsistent with a
%simple
%isotropic
%$s$-wave type symmetry of the order parameter
single $s$-wave order parameter
but is rather in favor of the presence of two gaps
without nodes.
%with the $s$-wave-like symmetry.
%or an anisotropic s-wave.
Our infrared reflection spectra supports the magnitude of the large gap, obtained from SH and lower critical field data,  and its nodeless character.
The IMARE spectroscopy data, obtained on SnS-Andreev arrays, refine the conclusions on the two nodeless gaps: the large gap,  $\Delta _{L} = 6.4 - 9.5$\,meV with extended $s$-wave symmetry and anisotropy in the $k$-space not less than $\approx$ 30\%,
and the small gap, $\Delta _S = 1.8 \pm 0.3$\,meV.
The BCS-ratio for the the upper extremum of the large gap is $2\Delta _L/k_BT_c \approx 6.1 - 7.2$. All our data clearly show that the superconducting energy gaps in nearly optimally doped Ba$_{1-x}$K$_x$Fe$_{2}$As$_{2}$ are nodeless.
%}
In addition, the obtained gaps
%from our measurements
are consistent with those determined from ARPES measurements.

%%%%%%%%%%%%%%%%%%%%%%%%%%%%%%%%%%%%%%%%%%%%%%%%%%%%%%%%%%%%%%%%%%%%%%%%%%%%%%%%%%%%%%%%%%%

\section{Acknowledgements}
\begin{acknowledgments}
%The authors thank ..... for fruitful discussions.
This work is supported by the Russian Science Foundation (16-12-10507).
Magnetic measurements were carried out with the support of the Russian Foundation for Basic Research (16-32-00663).
M.A. acknowledges funding by DFG in the project MO 3014/1-1.
YuAA acknowledges the support of the Competitiveness Program of NRNU MEPhI.
Authors also acknowledge the Shared Facility Center at LPI for using their equipment.
\end{acknowledgments}

\newpage
\section{Appendix 1}
\subsection{Data processing protocol}

As mentioned in the main text,  the noise level in our $M(H)$ measurements was rather high, $3-5 \times10^{-5}$\,emu with the signal of the order of  $10^{-2}$\,emu. We  firstly investigated the possibility of determination $H_{c1}$ using the conventional method  \cite{hafiez-FeSe_PRB_2013, hafiez-122_PRB_2014}. For this purpose we model the typical $M(H)$ dependence using a piecewise analytical formulae.
%For the analysis
We break the range of measurements  in two regions. In the low-field region we model the $M(H)$ data with a linear model dependence, whereas above a certain field $H_0$ -- with a parabolic one. The parameters for both linear and non-linear parts are fitted to the measured $M(H)$ data; the step size for the model dependences was chosen 0.5\,Oe.
We further
%take into consideration a random noise, by
add a random signal within a chosen noise level to each data point of the model $M(H)$ dependence, and consider how this noise affects the correlation coefficient $R$, calculated by the conventional method \cite{hafiez-FeSe_PRB_2013, hafiez-122_PRB_2014}, and also the extracted $H_{c1}$  value.

\begin{figure}%[h]
\begin{center}
\includegraphics[width=240pt]{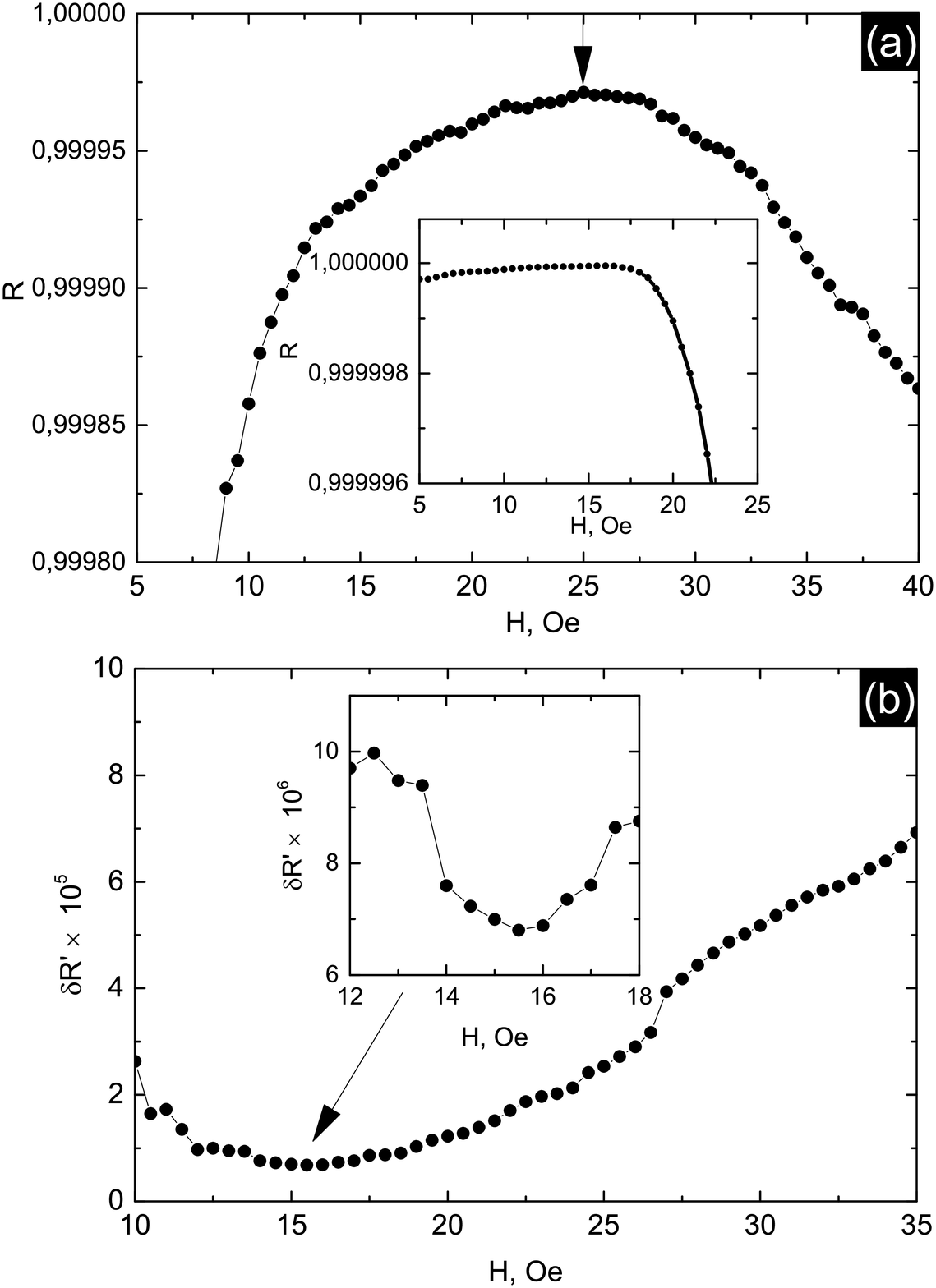}
\caption{(a) Correlation coefficient $R$ calculated for model data with noise level about $5 \times 10^{-5}$\,emu by method used in \cite{hafiez-FeSe_PRB_2013, hafiez-122_PRB_2014}.
Inset: correlation coefficient $R$ calculated for model data with noise level $10^{-6}$\,emu.
(b)  Deviation  of the correlation index from unity $\delta R'=1-R^\prime$ versus $H$ calculated by the modified method for the same model data.
}
\label{Fig:R1R2}
\end{center}
\end{figure}

It appears that the correlation coefficient calculated for the noise level about $10^{-6}$\,emu coincides with that calculated in Ref.~\cite{hafiez-122_PRB_2014}.
%(see inset in Fig.~\ref{Fig:R1R2}a).
Particularly, it exhibits a plateau below $H_{c1}$.
By taking the field value where $R$ starts sharply decreasing  we obtain the $H_{c1}$ value that also coincides with the $H_{0}$ parameter of the model.
The inset to Fig.~\ref{Fig:R1R2}\,a shows a typical correlation coefficient $R(H)$ calculated for the noise level $10^{-6}$\,emu. This dependence has a weak maximum at $H=16$\,Oe, which is only by 6\% higher than the parameter $H_0=15$\,Oe included in the model.

However, as noise increases, the $R(H)$ dependence changes drastically: the maximum becomes more clearly pronounced and its departure from $H_0$ increases. Figure \ref{Fig:R1R2}\,a shows the correlation coefficient $R$ calculated for the model $M(H)$ dependence with a $50\times$ bigger noise level,  $5\times 10^{-5}$ (typical for the experiment), and for $H_0=15$\,Oe. Instead of plateau, $R(H)$  here exhibits a maximum at $\approx 25$\,Oe which is essentially higher than the given $H_0$ value.

In order to overcome the problem of extraction the $H_{c1}$ value in the presence of noise, we have modified the above algorithm of Refs.~\cite{hafiez-FeSe_PRB_2013, hafiez-122_PRB_2014}.
%In fields above $H_{c1}$ the superconductor traps magnetic flux that leads to the departure of $M(H)$  dependence from the linear one.
In the modified method we expand the trapped magnetization $M$   as $M (H) \propto (H -H_{c1})^2$  in the vicinity of $H_{c1}$. Correspondingly,  the magnetization  may be written as follows:

\begin{eqnarray}
\label{ModelHc1}
M(H)&=& aH+b \nonumber \qquad\qquad\qquad\quad \text{\rm for } H<H^*\\
M(H)&=& aH+b +c (H-H^*)^2 \quad \text{\rm for } H>H^*.
\end{eqnarray}

%For all magnetic fields $H^*$ within the interval of our measurements,
%for which the magnetization has been measured,
%we fitted the M(H) data using $a, b$ and $c$ as fitting parameters (the parameter $b$ corresponds to an insignificant, an order of $10^{-4}$\,emu, possible residual zero field magnetization $M (H = 0)$ introduced for the best fitting of the experimental data with the model curve Eq. (1). Further, the regression coefficient $R(H)$ was calculated for the function $F (H) = H + M_{exper}/M_{fit}$  where $M_{fit}$ - is the magnetization calculated with the models with a given set of parameters a, b and c  (for a function f (x) = const, the regression coefficient is not defined, and requires transforming from the constant to a linear dependence by addition $H$).

For every running data point $H_i$ we take
%we took all points $H$ for which magnetization was measured as
$H^*=H_i$
and find the best fitting of the experimental data with the model curve Eq. (\ref{ModelHc1}), using
%and determined
$a, b$ and $c$ as fitting parameters (the parameter $b$ corresponds to an insignificant, an order of $10^{-4}$\,emu, possible residual zero field magnetization $M (H = 0)$).
%In order to characterize the accordance of experimental and the model
For every $H^*=H_i$  we calculate the
%correlation coefficient $R'(H)$
 correlation index
(coefficient of determination)
%that characterizes the
%accordance of the experimental and model data  was calculated
as follows:
\begin{equation}
R'(H)=\sqrt{1-\frac{\sum{(M_{exper}(H)-M_{fit}(H))^2}}{\sum{(M_{exper}(H)-\overline{M})^2}}}.
\end{equation}
Here $M_{fit}$ is the magnetization calculated within the model Eq.~(\ref{ModelHc1}) for the given set of parameters $a_i, b_i$ and $c_i$
 which are determined at point $H_i$, and $\overline{M}$ is the averaged magnetization value.
%(for a function $F (H) = M_{exper}/M_{fit}$, which is close to unity the correlation coefficient is not defined, and it requires transforming $F(H)$ from the constant to a linear dependence by addition $H$).
The model Eq.~(\ref{ModelHc1}) is expected to give the best fit of the experimental data at the  $H_i=H_{c1}$, 	
%since $H_{c1}$ is that point where dependence M(H) changes from linear to nonlinear
therefore we interpret the $R'$  maximum point as $H_{c1}$.

Figure~\ref{Fig:R1R2}\,b shows the deviation from unity of correlation index calculated using the modified method for the same model function $M(H)$ as that  used above for calculations of $R$ in Fig.~\ref{Fig:R1R2}\,a, and for the same noise level $5\times10^{-5}$\,emu. This dependence has a  maximum at $H=15.5$\,Oe  that agrees with $H_0=15$\,Oe used in the model $M(H)$.
By  random varying
%variation of the values of
the magnetization within the the same noise level we found that the maximum of $R'$ (and therefore $H_{c1}$) varies within 2\,Oe; we consider this as the estimate of the  uncertainty of $H_{c1}$.

\begin{figure}
\begin{center}
\includegraphics[width=240pt]{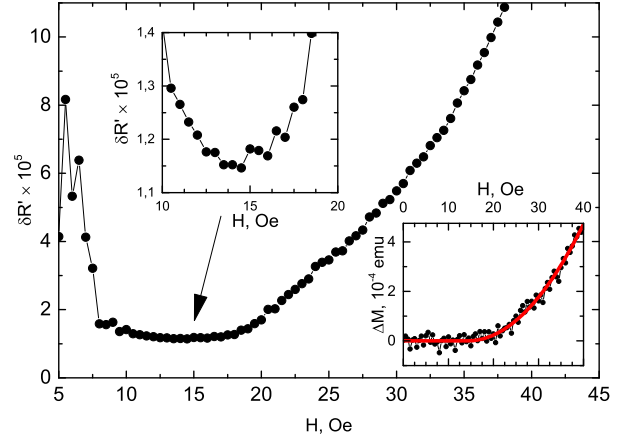}
\caption{ Deviation from unity of the correlation index  $\delta R'=1-R^\prime(H)$ calculated by the modified method  for the data taken at $T = 15$\,K. The upper inset shows the $\delta R'(H)$ dependence near its minimum. The lower inset shows  deviation of the measured $M(H)$ data from the linear $M(H)$ dependence.}
\label{Fig:Rnew}
\end{center}
\end{figure}

Figure~\ref{Fig:Rnew} shows the deviation from unity of correlation index $\delta R'=1-R'$,  versus $H$ calculated  from our experimental  $M(H)$ data measured at $T=15$\,K. The upper inset shows this dependence near it's minimum. This minimum is taken as the best estimate of $H_{c1}$. The lower inset shows deviation of the experimental  $M(H)$ data from the best linear fit calculated with parameters $a_i$ and $b_i$ defined for the point of maximum $R'(H_i)$.  The high fit quality demonstrates the applicability of the model Eq.~(\ref{ModelHc1}) to the experimental data.
% non-linear part of fitting by Eq. \ref{ModelHc1} to it.

\end{document}